# An improved understanding of the roles of atomic processes and power balance in divertor target ion current loss during detachment

K. Verhaegh[1,2], B. Lipschultz[1], B.P. Duval[2], O. Février[2], A. Fil[1,3], C. Theiler[2], M. Wensing[2], C. Bowman[1,3], D.S. Gahle[4,3], J.R. Harrison[3], B. Labit[2], C. Marini[2], R. Maurizio[2], H. de Oliveira[2], H. Reimerdes[2], U. Sheikh[2], C.K. Tsui[5,2], N. Vianello[6], W.A.J. Vijvers[7], the TCV team[8] and the EUROfusion MST1 team[9]

[1] *York Plasma Institute, University of York, United Kingdom*
[2] *Ecole Polytechnique Fédérale de Lausanne (EPFL), Swiss Plasma Center (SPC), CH-1015 Lausanne, Switzerland*
[3] *Culham Centre for Fusion Energy, Culham Science Centre, OX14 3DB, UK*
[4] *Department of Physics SUPA, University of Strathclyde, Glasgow, G4 0NG, UK*
[5] *University of California San Diego (UCSD), San Diego, United States*
[6] *Consorzio RFX, Padova, Italy*
[7] *DIFFER, Eindhoven, The Netherlands*
[8] *See the author list of S. Coda et al. 2019 Nucl. Fusion, accepted*
[9] *See the author list of B. Labit et al. 2019 Nucl. Fusion, accepted*

## Abstract

The process of divertor detachment, whereby heat and particle fluxes to divertor surfaces are strongly diminished, is required to reduce heat loading and erosion in a magnetic fusion reactor to acceptable levels. In this paper the physics leading to the decrease of the total divertor ion current ($I_t$), or 'roll-over', is experimentally explored on the TCV tokamak through characterization of the location, magnitude and role of the various divertor ion sinks and sources including a complete analysis of particle and power balance. These first measurements of the profiles of divertor ionisation and hydrogenic radiation along the divertor leg are enabled through novel spectroscopic techniques.

Over a range in TCV plasma conditions (plasma current and electron density, with/without impurity-seeding) the $I_t$ roll-over is ascribed to a drop in the divertor ion source; recombination remains small or negligible farther into the detachment process. The ion source reduction is driven by both a reduction in the power available for ionization, $P_{recl}$, and concurrent increase in the energy required per ionisation, $E_{ion}$: This effect of power available on the ionization source is often described as 'power starvation' (or 'power limitation'). The detachment threshold is found experimentally (in agreement with analytic model predictions) to be $\sim P_{recl}/I_t E_{ion} \sim 2$, corresponding to a target electron temperature, $T_t \sim E_{ion}/\gamma$ where $\gamma$ is the sheath transmission coefficient. The target pressure reduction, required to reduce the target ion current, is driven both by volumetric momentum loss as well as upstream pressure loss.

The measured evolution through detachment of the divertor profile of various ion sources/sinks as well as power losses are quantitatively reproduced through full 2D SOLPS modelling through the detachment process as the upstream density is varied.

## 1. Introduction

Divertor detachment is predicted to be crucial for handling the power exhaust of future fusion devices such as ITER [3]. Aside from target power deposition due to radiation and neutrals, the plasma heat flux ($q_t$ in W/m²) is dependent on the divertor target ion flux density ($\Gamma_t$ in ions/m²) and electron temperature ($T_t$ in eV) where $\gamma$ is the sheath transmission coefficient and $\epsilon$ is the potential energy deposited on the target per ion.

$$q_t = \Gamma_t(\gamma T_t + \epsilon) \tag{1}$$

The ion target current is constrained by the sheath as expressed in equation 2, where $p_t$ is the target pressure and a target Mach number of one is assumed.

$$\Gamma_t \propto n_t C_s \propto p_t/T_t^{1/2} \tag{2}$$

Detachment is a state in which the divertor plasma undergoes a large range of different atomic and molecular reactions to ultimately provide a simultaneous reduction of the ion target current and the target temperature and thus pressure, providing access to large target heat flux reductions (equation 1). In fact, a simultaneous reduction of the ion target current and the target temperature implies the target pressure must drop faster than $T_t^{1/2}$ (equation 2): $\frac{\partial \Gamma_t}{\partial T_t} < 0 \rightarrow \frac{\partial}{\partial T_t}(p_t/T_t^{1/2}) < 0$. A reduction of the ion current, while the divertor plasma is cooled through core density or impurity seeding ramps, is experimentally often used as an indicator for detachment [3-5]. Investigating the nature of the ion target current reduction is thus crucial for the understanding of detachment.

Atomic/molecular processes during detachment result in reductions of particle, power and momentum (pressure), *enabling detachment*: a simultaneous reduction of target temperature and the target ion flux [4-8]. In this work we study the relation between particle/power/momentum balance and the target ion flux reduction experimentally by monitoring changes in power/particle/balance due to a range of atomic processes utilising visible spectroscopy.

A target pressure reduction (e.g. $\frac{\partial \Gamma_t}{\partial T_t} < 0 \rightarrow \frac{\partial}{\partial T_t}(p_t/T_t^{1/2}) < 0$ – equation 2) is enabled by volumetric momentum losses [4, 5, 9] and/or a reduction of the upstream pressure. Volumetric momentum losses are often attributed to the dominance of ion-neutral reactions (e.g. charge exchange and ion-molecule) over ionisation reactions at low temperatures ($T_e \lesssim$ 5-10 eV [4, 9]). However, upstream pressure reduction can also occur during detachment and has been observed on COMPASS [10].

Detachment [8, 11-15] requires power and particle losses in addition to target pressure loss. The ions reaching the target surfaces come off the target as neutrals and are ionised in the divertor, after which they reach the target again. This recycling process is generally considered the dominant ion source in the divertor [4, 15]. As the ionisation process costs energy ($E_{ion}$), the magnitude of ionisation is limited by the power flux entering the recycling region, $q_{recl}$. This is illustrated by equation 3 (derived in section 3.3), assuming an absence of recombination [8, 12, 15]. Such an analytic model [8, 12, 15] as well as fluid models [12, 16-18], show that once the power required for ionisation and the power entering the recycling region become comparable (e.g. $\frac{\gamma T_t}{E_{ion}} \ll 1$), the divertor ion source must drop, reducing the divertor ion target flux, which is referred to here as 'power limitation'.

$$\Gamma_t = \frac{q_{recl}}{E_{ion}} \times \frac{1}{1+\frac{\gamma T_t}{E_{ion}}} \tag{3}$$

While viewing target ion current roll-over during detachment alternatively through the viewpoints of pressure loss or as a competition between ion sources (ionisation) and sinks (recombination) may appear to describe detachment differently, they are, in fact, not mutually exclusive and all cited processes can/will occur [4, 9, 11].

Particle balance studies often focus on investigating volumetric electron-ion recombination, which can be an effective ion sink with adequate divertor temperatures (<1 eV) and densities (> $10^{20}$ m$^{-3}$).

Volumetric recombination has been predicted to play a central role in the target ion flux reduction [19-25] and is sometimes found, through quantitative analysis, to be significant in the reduction of the ion target flux [8, 25-30]. However, previous work on TCV [31], C-Mod [8] and JET [32] (latter two both $N_2$ seeding) has shown conditions where the volumetric recombination rate is insufficient for explaining the observed reduction in ion target current.

As described earlier, the target ion current reduction requires a reduction of the ion source and/or an increase in the ion sink (volumetric recombination) in *'high recycling'* conditions. We define *'high recycling'* as the divertor ion sources dominating over ion flows into/out of the divertor. Determining those ion sources requires experimental measurements of ionisation and its power losses. Quantitative measurements on the divertor ion source have not yet been performed, although experimental indications for power limitation are available (either from inferred ion sources [8], or from qualitative spectroscopic 'indicators' based on $D_\alpha$ [33]). Recent studies aim to provide quantitative information on ionisation during divertor detachment [1, 32, 34].

In this work we utilize new techniques for achieving these measurements (see [2]) which allow us to directly infer a) both the recombination sink/*the ionization source* and b) the total energy 'cost' per ionization, $E_{ion}$ per ionisation event as profiles along the outer divertor leg. The results of those studies (section 3) for L-mode TCV discharges quantitatively verify that the loss of target ion current is primarily due to an ionisation source loss for the TCV cases shown. Volumetric electron-ion recombination, as an ion sink, has a smaller (and sometimes negligible) effect even after the target ion current begins its roll-over. This loss of ion source coincides with the power entering the recyding region ($P_{recl}$) approaching the power required for ionisation ($P_{ion}$) - providing experimental evidence for 'power limitation'. Given the spectroscopic coverage (Figure 1), this provides an unprecedented view of the evolution of the ionization, recombination, electron density and impurity radiation profiles along the divertor leg during detachment. Those profiles and their evolution in time are in quantitative agreement with SOLPS simulations [35].

The above results are then compared with reduced analytic model predictions (based on [4, 8, 9, 11, 22]) in section 4. Several predictions of these analytic models ($I_t$, $T_t$, ....) are in quantitative agreement with the experimental results. Besides this, here are three important outcomes of this comparison:

1. Combining the Bohm sheath criteria with power/particle balance (equations 2, 3) results in quantitative analytic predictions for *the detachment onset* – where obeying equations 2 and 3 simultaneously *requires target pressure loss* (Appendix A.1): $P_{recl}$ ~ 2 $P_{ion}$, $T_t \sim E_{ion}/\gamma$ and $p_t/q_{recl} \sim \sqrt{\frac{m_i}{2E_{ion}\gamma}}$ (where $m_i$ is the ion mass). Our experimental measurements confirm those analytic predictions at the point in time we designate as detachment onset – namely when the ion target current starts to deviate from its (attached) linear increase.

2. We show the equivalence of approaching detachment from momentum balance (e.g. target pressure losses) and power limitation arguments from combining the Bohm sheath criteria with power/particle balance (section 4.2 – equation 21). This is supported with experimental measurements which show that both power loss (in fact power-limitation of the ion source) and volumetric momentum loss occur after the detachment onset. In addition, upstream pressure loss occurs during detachment, which is shown to be consistent with analytic modelling.

3. The $I_t \propto n_{eu}$ trend observed experimentally in TCV (where n$_{eu}$ is the upstream electron density) during attached conditions contrasts the often assumed $I_t \propto n_{eu}^2$ trend on which the Degree of Detachment (DoD) is based [3, 7, 24, 36-38]. The TCV observations are however supported with analytic predictions, when accounting for changes in the upstream temperature and divertor radiation. This illustrates deviations in upstream and divertor conditions need to be accounted for before the DoD can be used.

Our measurements show that as further power limitation occurs (P$_{recl}$ gets closer to P$_{ion}$), volumetric momentum loss (estimated from inferred charge exchange to ionisation ratios), molecule-plasma interaction (evident from an increase in $D\alpha$ [1]) and ultimately, volumetric recombination (P$_{recl}$ ~ P$_{ion}$), occur. This sequence, which is commenced and driven by power limitation appears to occur more generally in high recycling divertors and applies to various approaches to detachment – density scans, P$_{SOL}$ reductions and impurity seeding. It also applies to higher density experiments where the effect of recombination, while larger than for TCV, must still await the drop in the T$_t$ to low enough values (~ 1eV) driven by power limitation. Those low values occur after momentum loss starts to occur (~ 5eV) and roll-over starts (~ 2 eV in TCV).

## 2. Experimental setup

All the research discussed involved L-mode Ohmic density/impurity ramp discharges made in the medium-sized tokamak TCV (R = 0.89 m, a = 0.25 m, B$_t$ = 1.4 T)[39]. The characteristics of the various discharges utilised, as well as their equilibria, are shown in table 1 and Figure 1a, respectively. All discharges are in L-mode without additional heating and are performed in reversed field (e.g. $\nabla$B in the unfavourable direction) to stay out of H-mode. These choices have been made in order to obtain detached conditions with relatively high divertor densities (n$_e$ ~ 10$^{20}$ m$^{-3}$), which so far have not yet been achievable in H-mode or heated L-mode discharges [40]. Reversed field is required to obtain such densities as otherwise the plasma current would need to be reduced to stay in L-mode [24, 37, 38]; which would imply reductions in core, upstream and divertor densities. Expected deviations between our TCV results and other devices; H-mode and forward-field conditions are discussed in section 4.4 and are expected not to influence the main conclusions of this work.

To obtain ionisation sources and sinks, we utilised the newly developed TCV divertor spectroscopy system (DSS) [1, 2, 31]. The DSS consists of vertical and horizontal viewing systems, each employing 32 lines of sight (Figure 1a). *Our analysis is based on the horizontal system*, which provides full coverage for the divertor shapes studied in this work. Full details on the analysis can be found in [1, 2] and a summary can be found in section 2.1.

Other diagnostics used for portions of the work presented are gold foil bolometers, target Langmuir probes [41], an upgraded Thomson scattering system [42], a reciprocating probe [43] and infrared imaging [44]. The locations of these different diagnostics are shown in Figure 1b.

We have divided the radiated power into core radiation (above the x-point) and divertor radiation (below the x-point). This is accomplished by utilising the brightness from poloidal bolometric chords over the appropriate region, while removing chords which intersect the inner divertor (to prevent contamination from inner divertor radiation). Such an analysis of bolometric chordal brightnesses has been used in place of the 'default' tomographic reconstruction of the radiated power emissivity across the entire plasma which can have significant uncertainties [40, 45, 46]. We note that due to the reflection of low energy photons from the gold foil of the bolometers, the estimated radiated power is assumed to be underestimated by at least 15% [47]. When considering other uncertainties, the overall underestimate of the radiated power ranges from 10% - 20%, which we correct for in our analysis. Further explanation on the bolometric analysis can be found in [1].

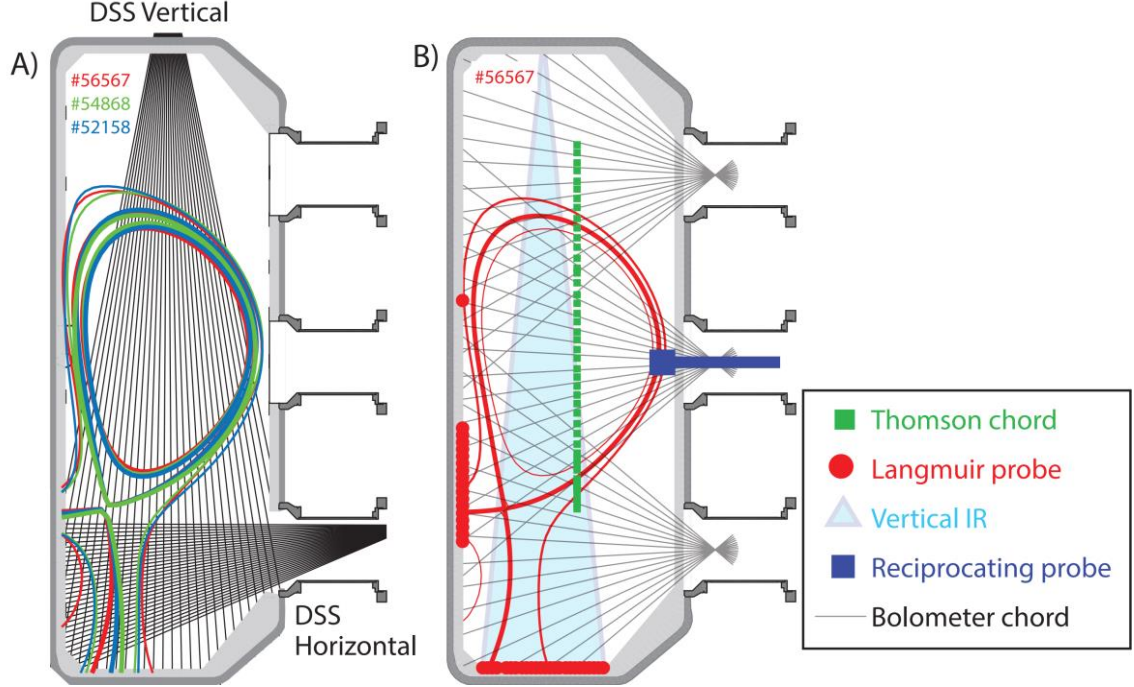

*Figure 1. a): Lines of sight of the horizontal and vertical DSS systems. Divertor geometries for #56567 (red), #54868 (green), #52158 (blue) are shown. b) Lines of sight and locations of other diagnostics (Thomson/Langmuir probes/Vertical IR/Reciprocating probe/Bolometry together with the divertor geometry of #56567.*

| Discharge number | $I_p$ (kA) | Greenwald fraction | |
|---|---|---|---|
| 56567 (and repeats) | 340 | 0.3 – 0.6 | Density ramp |
| 54868 | 240 | 0.25 – 0.6 | Density ramp |
| 52158 | 340 | 0.4 | $N_2$ seeding ramp |
| 57912 | 340 | 0.25 – 0.6 | Density ramp |

*Table 1: Overview of discharges used in this work, together with their shot numbers, plasma current and Greenwald fraction.*

The values of power entering the recycling region, $P_{recl}$, can be sensitive to this underestimation during detached conditions where $P_{recl}$ is small. This is reflected in our results, as we assume that 80-90% of all radiative losses is detected by the bolometry, for which we assume a uniform probability density function in the probabilistic analysis used for quantities which require $P_{recl}$ (or $q_{recl}$) estimates [2]; which includes the analytic model investigations in chapter 4.

## 2.1 Spectroscopic analysis methodology

We first provide a brief summary of our analysis techniques and nomenclature for inferring the recombination sink [31], ionisation source and hydrogenic power loss. The manuscript on this analysis technique can be found at [2], while its code is available at [48]. Figure 2 illustrates the various steps in the analysis process, eventually resulting in estimates of both local plasma characteristics (weighted over the Balmer line emission profile along each viewing chord) and line integrated plasma parameters. This analysis strategy contains three steps.

1. The Balmer line shape is analysed to obtain an estimate of the characteristic electron density of the Balmer line emission region through Stark fitting the Stark broadened component [1, 2, 31].

2. The ratio between two Balmer lines is used to separate the excitation/recombination contributions of the Balmer line emission quantitatively ($B_{n\to2}^{rec}$, $B_{n\to2}^{exc}$ in photons / m$^2$ s, respectively) [1, 2, 31] by inferring the fraction of the Balmer line brightness due to recombination, $F_{rec}(n)$, and excitation, $F_{exc}(n)$ and multiplying those fractions with the absolute Balmer line intensity ($B_{n\to2}$ in photons / m$^2$ s).
3. $B_{n\to2}^{rec}$ and $B_{n\to2}^{exc}$ are analysed individually to infer chordally-integrated parameters, such as the recombination ($R_L$ in rec/m$^2$s) rate, ionization ($I_L$ in ion/m$^2$s) rate and radiative power loss due to excitation and recombination ($P_{rad,L}^{exc}$, $P_{rad,L}^{rec}$ in W/m$^2$, respectively) [2]; as well as a measure of the local (along a chord) 'characteristic' excitation/recombination temperature ($T_e^E$, $T_e^R$, respectively). Those can be interpreted as emission-weighted temperatures along the line of sight [2]. Using the excitation temperature $T_e^E$ and assuming that excitation and charge exchange occur at the same location of the chordal integral, an estimate of the line integrated charge exchange to ionisation ratio $CX_L/I_L$ can be obtained.

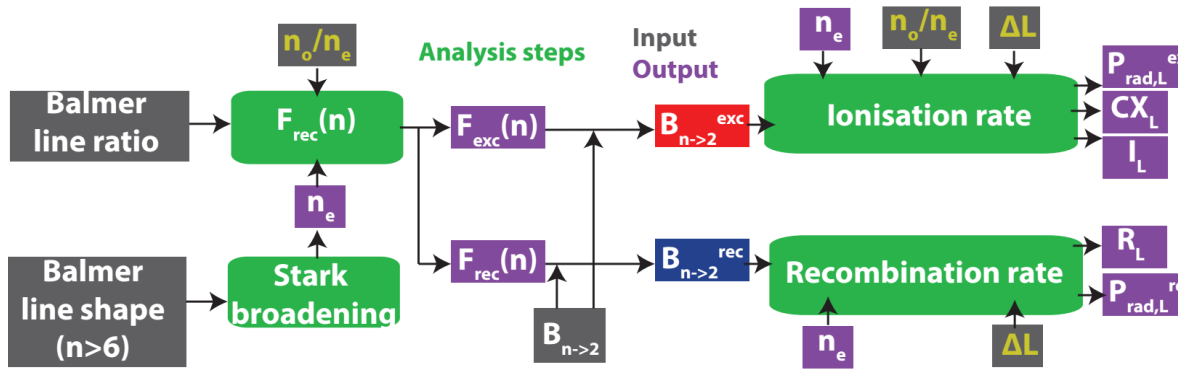

*Figure 2: Schematic overview of recombination and ionisation rate analysis methodology. Inputs are shaded in grey, assumed inputs have yellow symbols, outputs in purple, analysis steps in green.*

The chordal-integrated parameters can then, provided one has a full coverage of the divertor, be toroidally/poloially integrated (using the technique in [2, 27]) to provide the total divertor recombination rate $I_r$ (in rec/s); ionisation rate $I_i$ (in ion/s); hydrogenic excitation radiated power $P_{rad}^{exc}$ in W; hydrogenic recombinative radiated power $P_{rad}^{rec}$ in W. We have performed this analysis for the outer divertor leg. Further information on all the various output parameters can be found in table 2 in [2].

As shown in Figure 2, several input parameters (e.g. neutral fraction $n_o/n_e$, pathlength $\Delta L$) are required and assumptions must be made to characterize them – see table 1 in [2] for an overview. The uncertainty for some of those parameters can be too large for Taylor-expansion based error analysis techniques to be appropriate. We thus developed and used a Monte-Carlo based probabilistic analysis [2] to estimate all output quantities and their uncertainties. This works by ascribing a probability density function (PDF) to every single input parameter in the analysis, characteristic of their uncertainties. Random values are sampled according to those PDFs on which the analysis is performed. This, eventually, leads to a distribution of different output parameters which is mapped to a PDF [49]. Using those output PDFs, an estimate of the output value and its uncertainty (Highest Density Interval) are obtained using techniques adopted from Bayesian analysis [50] - (Maximum Likelihood / Highest Density Interval [51]). See [2] for a full overview of this probabilistic technique and for examples of output PDFs.

The above analysis is based on a Balmer line slab model for the Balmer line brightness ($B_{n\to2}$ in ph/m$^2$ s with upper quantum number $n$) - Equation 4, which models the Balmer line emission as if it originates from a plasma slab with spatially constant parameters (0D model) with a chord intersection length of $\Delta L$. $B_{n\rightarrow 2}$ is a function of path length ($\Delta L$), electron density ($n_e$), neutral density ($n_o$) and temperature ($T_e$) using the Photon Emissivity Coefficients ($PEC_{n\rightarrow 2}^{rec}$) for recombination and excitation ($PEC_{n\rightarrow 2}^{exc}$), obtained from the Open-ADAS database [52, 53].

$$B_{n\rightarrow 2} = \underbrace{\Delta L\, n_e^2\, PEC_{n\rightarrow 2}^{rec}(n_e, T_e)}_{B_{n\rightarrow 2}^{rec}} + \underbrace{\Delta L\, n_e n_o PEC_{n\rightarrow 2}^{exc}(n_e, T_e)}_{B_{n\rightarrow 2}^{exc}} \quad (4)$$

Molecular reaction contributions (which can contribute strongly to $D_\alpha$ [1]) to the Balmer line emission are neglected in equation 4, which is valid for n>4 Balmer lines. Furthermore, equation 4 assumes the hydrogen ion density equals the electron density (e.g. $Z_{eff} = 1$) – which introduces insignificant errors on the analysis shown below [2, 31].

Deviations between the inferred parameters and actual parameters can occur due to the above assumptions as well as 'line integration effects': artefacts in the analysis output arising from the fact that the chord integrates through a plasma with spatial profiles rather than the 0D slab model presented in Equation 4. This has been investigated in detail in [2] – indicating that the analysis is insensitive to the mentioned assumptions as well as line integration effects. The insensitivity is determined by using a synthetic diagnostic approach applied to SOLPS simulations [2] as well as by using more simplified a priori model with assumed $n_e$, $T_e$ profiles [31]; deviations between inferred parameters and actual parameters remain smaller than the characteristic uncertainties of the inferred parameters [2]. By incorporating 2D spectroscopic measurements using filtered camera imaging [54], this analysis may be further improved.

## 2.2 Reproducibility of repeat discharges

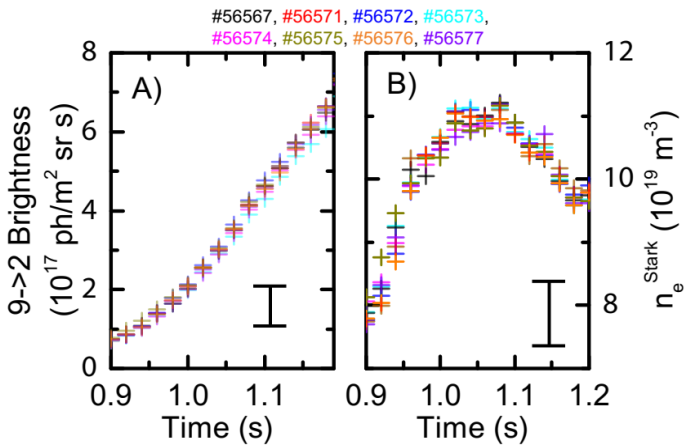

Figure 3: a) 9->2 Balmer line brightness and b) inferred Stark density from the 9->2 Balmer line obtained from the vertical system using the line of sight closest to the strike point location. Each colour indicates a different discharge. Characteristic uncertainties are shown in the Figure.

At sufficiently high electron densities and recombination rates, a lower-n Balmer line (n=5) is required for separating excitation/recombination emission quantitatively. However, a higher-n Balmer line (n-7) is required for the Stark density inference [2]. To facilitate both measurements, diagnostic repeats are required, and thus the reproducibility of repeat discharges must be verified. To demonstrate such reproducibility we show, in Figure 3, the variation of the brightness and Stark density measurements for a set of 8 sequential identical discharges (of which #56567 is studied extensively in this paper) using the vertical DSS spectrometer from the line of sight corresponding to the strike point. The time dependencies of Balmer n=9 line intensity (Figure 3a) and the derived chordal averaged (weighted by the n=9 (recombinative) emission profile) density (Figure 3b) are the same within uncertainty from discharge to discharge. In addition, results from other diagnostics (bolometry and Langmuir probes – not shown) also agree within uncertainty for the repeated discharges, indicating enough reproducibility for our primary measurements of the divertor plasma characteristics. The reproducibility can be significantly worse if discharges are repeated on different days.

# 3. Results

Particle balance measurements show the ionization source is the primary process that determines the target ion current and its reduction during detachment in TCV. Volumetric recombination plays either a secondary or negligible role in the ion current reduction and only occurs *after* the ionisation source rolls-over. These results have been obtained for three discharges: two core density ramps (at two different plasma currents) and a nitrogen seeding ramp.

Power balance measurements shows 'power limitation' reduces the ion source, in agreement with theoretical predictions [15, 22]. The power reaching the recycling region, $P_{recl}$, is *reduced* during a density ramp discharge due to increasing impurity radiation; which is inferred from hydrogenic and total radiation estimates. In contrast, the power 'required for ionization', $P_{ion}$, *increases* during the pulse until $P_{ion} \sim P_{recl}$ at the target ion current roll-over. This provides experimental evidence for power limitation, which is concurrent with the development of pressure and momentum losses (section 4).

## 3.1 Detachment characteristics on TCV

First, we characterize the development of ion target current loss during detachment of a density ramp discharge. Secondly, we describe the development of the poloidal profiles of plasma characteristics during detachment experimentally, which are also compared with equivalent synthetic measurements [2] performed on the corresponding SOLPS plasma solutions [35]. Previous studies have provided complementary descriptions of the development of detachment in TCV [24, 31, 38, 46, 55] and are thus useful for further details.

### 3.1.1 Characterization of target ion loss

Feedback control of the $D_2$ fuelling was used for #56567 (Figure 4) to obtain a linear increase (with time) of the line averaged core density, $\overline{n_e}$, measured by a vertical FIR interferometer chord. This causes a linear increase of the upstream separatrix density, $n_{eu}$ (Figure 4). $\overline{n_e}$ is increased until the plasma disrupts at t=1.25 s, achieving a maximum Greenwald fraction of ~ 0.65. Both the total ion target flux integrated across the divertor target ($I_t$ in ion/s) and the target ion flux density at the separatrix ($\Gamma_t$ in ion/m$^2$ s) increase linearly initially (Figure 4a). The linear scaling of $I_t$ and $\Gamma_t$ with the upstream/core density for attached plasmas was observed for all the density ramp studies at TCV [38]. This contrasts the $\Gamma_t \propto n_{eu}^2$ scaling often *assumed* in other tokamaks [7, 36]. From a theoretical perspective, the $\Gamma_t \propto n_{eu}^2$ scaling assumes that the upstream temperature and target heat flux is unaltered as the upstream density is increased (chapter 2,3 [1]; [4]; [3]); which is not necessarily true. Accounting for measured changes in these parameters over the discharge, we will show in section 4.2.1 that the observed linear increase of $I_t$ and $\Gamma_t$ with $n_{eu}$ is expected when considering the measured reductions of upstream temperature and the increase in divertor radiation throughout the discharge as $n_{eu}$ is increased.

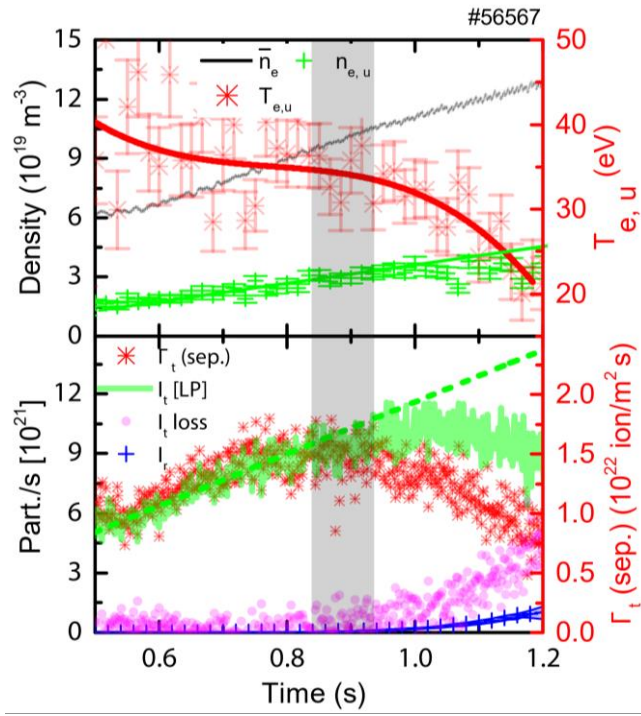

*Figure 4: Overview of detachment based on a 340 kA, density ramp discharge (#56567). a) Line averaged, $\overline{n_e}$, upstream density, $n_{e,u}$ and upstream temperature, $T_{e,u}$ as function of time. b) Total Ion target flux ($I_t$), ion target flux density at the separatrix ($\Gamma_t$), recombination rate ($I_r$) and $I_t$ loss as function of time. The onset of detachment phase (~0.82-0.87 s) is indicated as a black shaded region in this Figure and all subsequent experimental Figures as function of time.*

To quantify the loss of target ion current for this study we determine a linear, in upstream density and thus time, fit to the ion target current during the attached phase and extrapolate into the detached phase. The '$I_t$ loss' is then the difference from this to the measured ion target current, $I_t$ (see Figure 4b). At ~0.82-0.87 s $I_t$ starts to deviate from its linear trend and ion target current loss starts to occur; *we use that deviation to define the onset of the process of detachment.* Later, we will show that this time is in accordance with analytic detachment onset predictions (section 4.2) and corresponds to when the ionisation profile peak lifts off the target (section 3.1.2). For the case shown in Figure 4, $\Gamma_t$ rolls-over at the detachment onset while $I_t$ rolls-over later. The separation between the detachment onset, $I_t$ roll-over (negative slope in $I_t$) and $\Gamma_t$ roll-over can, however, vary from one discharge to another (see Figure 7d-f and [38]). The $I_t$ roll-over is subsequent with a flattening/reduction of upstream density for the case shown in Figure 4.

Although spectroscopic signatures of recombination start to appear just before the ion target flux roll-over, the $I_t$ loss (magenta – Figure 4b) is significantly larger than the total recombination sink integrated over the entire outer leg ($I_r$ blue - Figure 4b), indicating recombination alone is insufficient (at least by a factor three) to fully explain the $I_t$ roll-over. This observation is general on TCV ([31] and section 3.2) and has also been observed under higher density conditions in Alcator C-Mod [8] as well as under $N_2$ seeded conditions [32].

### 3.1.2 Experimentally observed TCV detachment dynamics and corresponding SOLPS solutions

The experimentally-measured poloidal profiles of several plasma parameters along the outer divertor leg during the periods before, during and after the target ion current roll over are compared to SOLPS simulation results (Figure 5). The profile times correspond to the vertical lines in Figure 5a. The equivalent density scan modelled using SOLPS is shown in Figure 5b. This simulation [35] does not reproduce the experimental result that the upstream density saturates upon detachment. As such, a linear trend of the upstream density has been used to match the chosen times to the appropriate $n_{eu}$.

The three SOLPS simulations used to compare to experimental profiles are indicated by the enlarged symbols in Figure 5b. Their colours correspond to the vertical lines at which the experimental data is taken, shown in Figure 5a. The SOLPS profile results (Figure 5f, h, j) are obtained by integrating through the 2D SOLPS profiles of ionisation, recombination, etc.) along the DSS and bolometric viewing chords [1, 2] (Figure 5h - $P_{rad,L}$), enabling a closer comparison between experiment and simulation. The divertor-integrated results (Figure 5f) are obtained by integrating the ionisation source/recombination sink from SOLPS over the region covered by the entire horizontal DSS horizontal viewing chord fan (Figure 1a). The SOLPS 'Stark density' result (Figure 5d) for each viewing chord is obtained from a synthetic DSS diagnostic. Further details of how the synthetic measurements created from SOLPS output are provided in [1, 2].

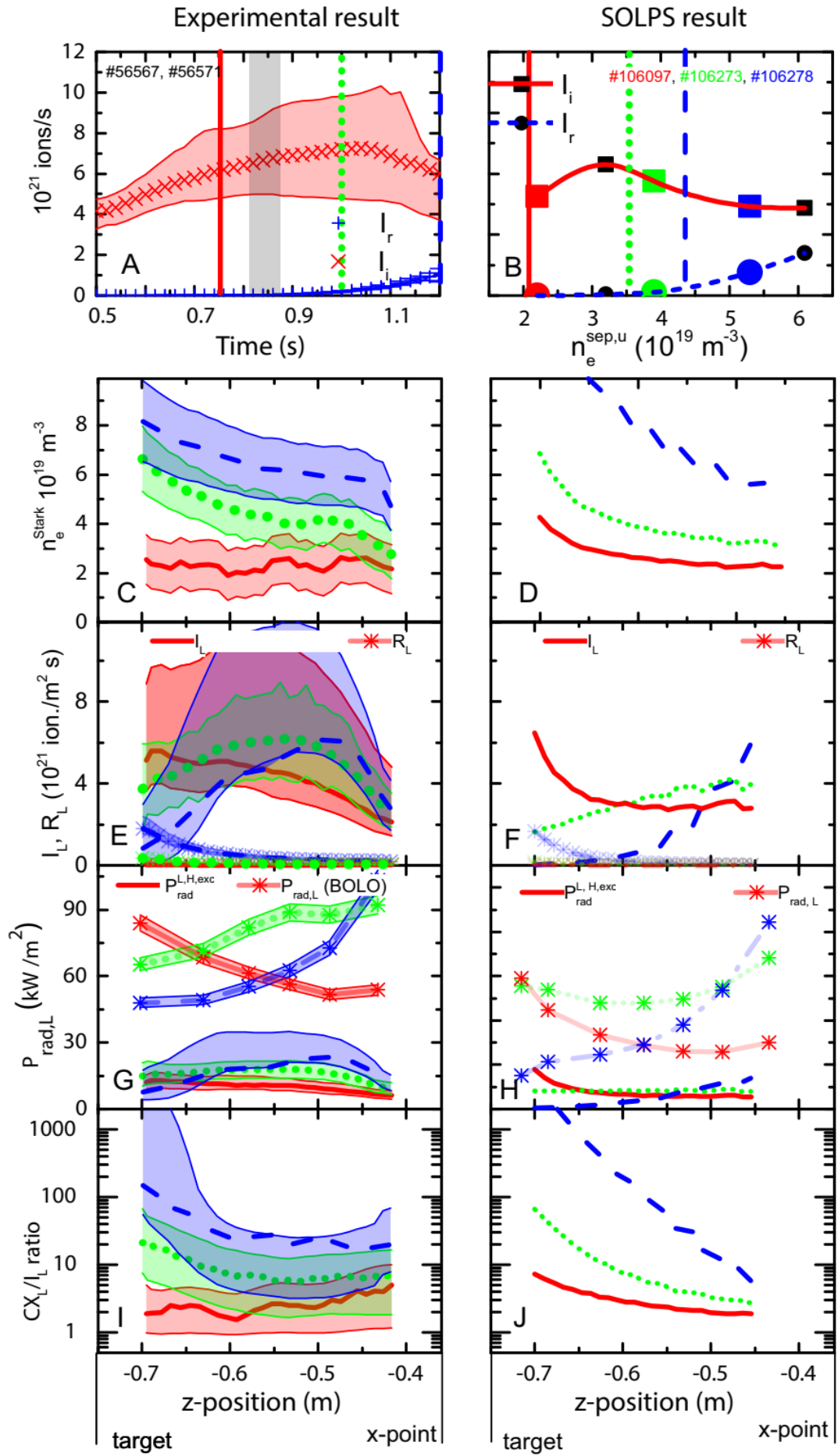

*Figure 5: Left hand side: Experimentally (spectroscopic inferences + bolometry) determined quantities along the outer divertor leg. Right hand side: Results obtained directly from SOLPS simulation utilizing synthetic diagnostic measurements. a) Outer divertor integrated ion source and recombination rate ($I_i$, $I_r$) as a function of time. The vertical lines correspond to the times at which the profiles are shown in the Figures below. b) Analogous ion source/sink plot (outer divertor integrated) obtained from SOLPS where the ionisation source and recombination sinks are shown as function of upstream density. The shot numbers in figure 5b are the reference numbers under which the SOLPS simulations are stored in the SOLPS MDSplus database. c, d) Stark density profiles (c – obtained from a synthetic diagnostic – see [1, 2].). e, f) Chordal integrated recombination ($R_L$) /ionisation rate ($I_L$) profiles. g, h) Chordal integrated total radiation profiles through bolometry – $P_{rad,L}$; and radiation due to hydrogenic excitation – $P_{rad,L}^{H,exc}$. I, j) Line integrated charge exchange ($CX_L$) to ionisation ratio ($I_L$) profiles.*

Before detachment, the density along the divertor leg (Figure 5c), the radiated power (Figure 5g – $P_{rad,L}$), ionisation rate (Figure 5e – $I_L$) and recombination rate (Figure 5e – $R_L$) all peak near the target. The ionisation region covers most of the divertor leg, which is likely due to the relatively large ionisation mean free path on TCV (5-10 cm). Further increases in $\overline{n_e}$ and $n_{e,u}$ generate a gradual shift of the radiated power peak towards the x-point, followed by the ionization region 'detaching' from the target (Figure 5e – $I_L$) concurrent with the detachment onset (starts at ~0.83s Figure 5a). As the ionization moves away from the target, a region where charge exchange dominates over ionisation is left behind (Figure 5i); eventually extending over a region up to ~ 20 cm from the target. During the entire detached phase, both the Stark density and recombination rate continue to increase across the entire divertor leg whilst their peaks remain near the target (Figure 5c,e) where the lowest DSS measurement chord is ~ 5 cm above the target surface. At the highest core density, recombination dominates over ionisation only over a small region (<10 cm) close to the target (Figure 5e).

Each chord passes through three regions of the outer divertor leg – a) the far SOL of the common flux regions (17 flux tube; 11.5 cm target coverage); b) the region near the separatrix (4 flux tubes; 0.5 cm target coverage); c) the private flux region (17 flux tubes; 13.4 cm target coverage). Comparing the total ionisation in these three regions of the SOLPS-ITER simulations. has shown that a negligible amount of ionisation (compared to the total) occurs at the private flux region (c) during both the attached (<4%) and detached phase (<1%). As the density ramp progresses the ionisation region widens across the far SOL of the common flux region (a) giving rise to more ionisation outside of the separatrix, increasing from 65% (attached) to 95% (detached) of the total ionisation.

All these observations are in excellent qualitative (and in most cases even quantitative) agreement with the SOLPS simulation results. However, certain parameters are different in the experiments and TCV SOLPS simulations [18, 35, 56]. In particular, the total ion target current trend in the simulations flattens during detachment as opposed to rolling over - in disagreement with the experiment. Although ionisation makes up most of the ion target current in the simulations in attached conditions, during detachment there are substantial ion flows into the divertor, balancing out the reduction of the total divertor ionisation source [18, 35, 56]. This is in disagreement with the measurements of section 3.2 for which there is quantitative agreement between the total divertor ion source and the total ion target current of the outer divertor leg in both the attached and detached phases, which is further discussed in section 4.3. The omission of drifts in the simulations could lead to this discrepancy of the ion flow into the divertor between the experiment and simulation. Those ion flows are, however, small for single flux tubes slightly outside the separatrix in the common flux region, where a clear roll-over (more consistent with the experiment) is recovered [18].

### 3.1.3 The dynamics of the electron density in the divertor during detachment

The three time points in the general plasma characteristic profiles along the divertor leg (Figure 5c, f, g, h) provide a coarse temporal resolution and therefore do not fully convey the dynamics of the electron density near the target, which we expect, based on previous work [28, 57, 58], to drop as the low pressure/density regions expand from the target towards the x-point during detachment.

Stark density measurements from the 7 horizontal DSS viewing chords closest to the target are shown in Figure 6a together with the viewing geometry (Figure 6b). This discharge is similar to the one discussed in sections 3.1.1 and 3.1.2, but with a magnetic geometry optimised for DSS strike point coverage. The inferred Stark densities and the target density measured by Langmuir probes both rise initially during the start of the density ramp. At approximately the time of the detachment onset and the total ion target current roll-over (~0.87s – not shown), which coincides with the time where the ion target flux deviates from its linear trend (not shown), the Stark densities throughout the divertor rise above the Langmuir probe target density, which rolls-over. Ultimately, the Stark density (within ~ 5 cm from the target) rolls-over (1.05 s). This consistent with observations from the vertical DSS system indicating a reduction in line averaged (9->2 Balmer line, thus recombination emission weighted) density throughout the divertor leg (Figure 3b).

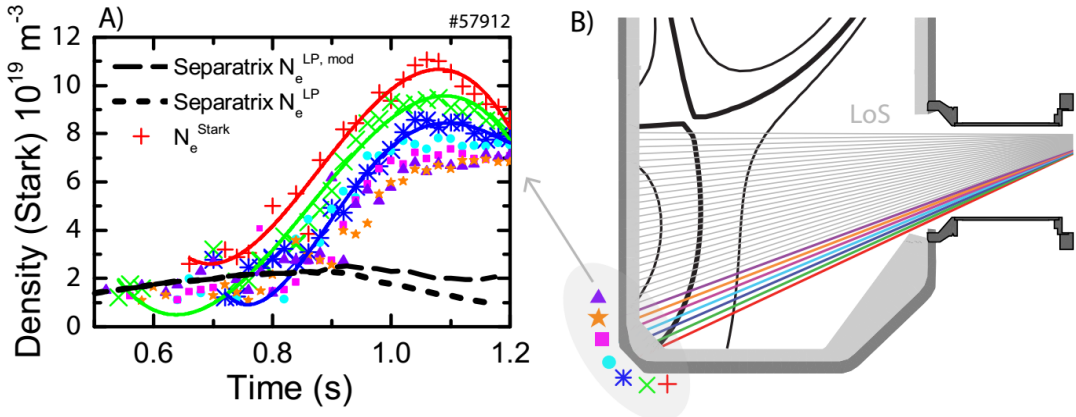

*Figure 6: A) Electron density (characteristic uncertainty ~$10^{19}$ m$^{-3}$) traces density from target Langmuir probes and DSS chords near the target (#57912) for a density ramp experiment. B) Divertor geometry and line of sights corresponding to the DSS measurements.*

The discrepancy between the Stark and Langmuir probe densities suggests that the electron density strongly decreases in a narrow region (< 5 cm) close to the target. The roll-over of the Stark density of the lowest viewing chord density would then be consistent with the density peak *starting to move* up along the leg. A decay of the electron density in such a narrow region during detachment is also observed in the SOLPS simulation (section 3.1.2) [35], although the amount that the electron density drops (1-2 . $10^{19}$m$^{-3}$) is significantly smaller than experimentally-inferred and shown in figure 6a.

However, there are other reasons which could partially explain the strong discrepancy between the Stark and Langmuir probe target densities.

1. There is a concern that the Langmuir probe measurement of the target density is incorrectly low. Since the Langmuir probe density inference uses the Langmuir probe-derived temperature ($J_{sat} \propto n_e^{LP}\sqrt{T_e^{LP}}$), the density would be underestimated when $T_e^{LP}$ is overestimated – which generally occurs in cold divertor conditions [41, 59, 60]. As such, following a similar approach as in [31], we calculated a modified $n_e^{LP,mod} = n_e^{LP}\sqrt{\frac{T_e^E}{T_e^{LP}}}$ using a spectroscopically inferred $T_e^E$ (section 4.1.1) from the excitation emission of the chord closest to the target (Figure 6a). $n_e^{LP, mod}$ remains significantly smaller than the observed Stark density upon detachment.

2. Some combination of the width of the poloidal viewing chord (1-2 cm) and the weighting of the Stark density towards the higher densities (and thus higher emissivities) could explain partially the difference between the target Langmuir probe density and the Stark density.
3. The Stark electron densities measured in low temperature conditions could be overestimated by up to $2.10^{19}$ m$^{-3}$ at the end of the discharge shown [1], due to the electron temperature dependence of the Stark width depending on the Stark model used [61].

## 3.2 Characterization of the loss of ion source and its effect on the ion target flux

In the survey of discharge characteristics (Figure 4), the inferred ionization source magnitude and time dependence appears to determine the current reaching the target. The following discussions are based on particle balance over the entire divertor and not just a particular flux tube. The balance of sources and sinks *within* the divertor can be written:

$$I_t \approx I_i - I_r \qquad (5)$$

where the target ion flux (the sink for ions at the target), $I_t$, is the sum over the divertor target surface while both the ionization source, $I_i$, and the volumetric recombination sink, $I_r$, are integrated over the entire outer divertor leg. Equation 5 assumes the divertor to be a closed, self-contained, system where the total divertor ion target current is dominated by divertor ion sources, ignoring sources of ions outside the divertor (core or SOL ionization) which flow from upstream towards the target; an approach used previously [8, 11, 12] and we will discuss it further in section 4.3. In this paper we define the divertor to be 'high recycling' when this condition (Equation 5) is valid.

### 3.2.1 Characterization of ion sinks and sources in density ramp discharges

We show examples of the equivalence of the divertor ionization source and target ion current in the first two columns of Figure 7 for density ramp discharges at two different plasma currents. The ionisation source (Figure 7d & e), $I_i$, tracks the increasing target flux, $I_t$, (within uncertainties) during the attached phase for both density ramp cases while recombination, $I_r$, is either negligible (Figure 7d) or small (Figure 7e). We conclude that the majority of ion target flux derives from ionisation within the divertor, in agreement with the self-contained divertor approximation (Equation 5), which shows that TCV is operating under 'high recycling' conditions. These measurements also indicate that any additional source of ion flux from the SOL into the divertor should be either relatively small or balanced by the ion flux flowing from the outer divertor towards the inner target.

High recycling divertor operation has been illustrated as a narrow ionisation region in front of the target [4]. This contrasts with our TCV observation (Figure 5). This indicates that having a narrow ionisation region may not be necessarily a requirement for cases where Equation 5 applies.

As explained in section 3.1.1, the ion losses are calculated by subtracting the measured $I_t$ and $I_i$ from these respective linear scalings of $I_t$ and $I_i$ in the attached phase (Figure 7d, e). The measured target ion current loss and the ionization source loss track well within uncertainties for both density ramp cases (Figures 7g, h). The recombination ion sink is only significant at the end of the high plasma current discharge; it only starts to develop to significant levels after the ion target flux roll-over and long after the deviation of the measured $I_t$ from its linear (attached) scaling and it remains more than a factor 4 lower than the loss of target ion current or loss of ionization source.

The recombination rate in the high current case is 5-10 times higher at the same core Greenwald fraction as in the lower current case. One explanation for the higher recombination rates is that the ~

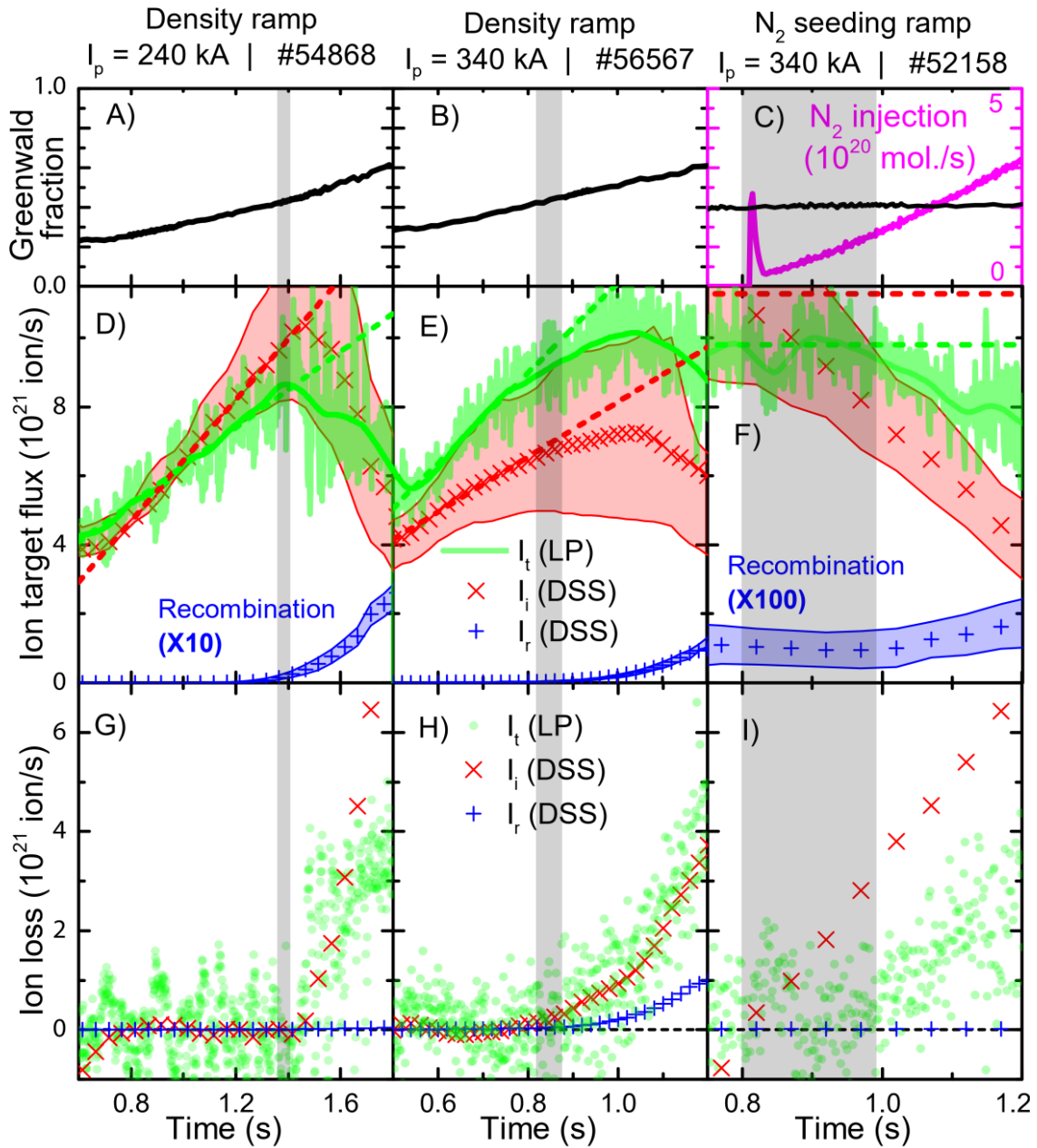

*Figure 7 – First two columns correspond to core density ramps at two different plasma currents: Core Greenwald fraction (a,b); divertor ion sources/sinks and ion target flux (d, e) as well as the loss of ion target current, recombination sink and loss of ionization (g, h). The last column corresponds to a $N_2$ seeding ramp at constant core density (c): divertor ion source/sink and ion target flux (f) as well as the loss of ion target current, recombination sink and loss of ionization (i).*

1.5 x higher $n_{eu}$ leads to higher divertor densities. These are observed (Stark density) to be ~ 3 x higher for the high current case, which agrees with the expected strong dependence of the divertor density on $n_{eu}$ (cubic, based on the two point model [4]). Assuming identical divertor temperatures between the two cases, this would result in ~10 times higher recombination rates (estimated from ADAS tables [52, 53]). This suggests that (for the same core Greenwald fraction) the plasma current is a 'control knob' for the influence of recombination on the ion target flux.

### 3.2.2 Characterization of ion sinks and sources in N$_2$-seeded discharges

N$_2$ seeded discharges develop significantly differently than the core density ramp discharges discussed previously [62]. The line averaged density for this pulse (Figure 7c) is held constant over the N$_2$-seeding ramp at a Greenwald fraction of ~0.4. This is just below the core density at which the ion target flux roll-over occurred in the equivalent high current, density ramp discharge (pulse #56567; Figure 7 b,e). The ion target loss is quantified as previously using the pre-N$_2$ seeding scaling as a reference. This likely underestimates the actual value of the ion target loss as the ion target flux is expected to increase with increasing impurity radiation in attached conditions while other divertor parameters are kept constant (Appendix A.1 - Equation A.5 (or A.6) together with the target temperature solution of equation A.7). The magnitude of the ion source loss, including the range of uncertainty, was larger than that needed to explain the magnitude of the ion target flux.

We cafn only speculate as to why the particle balance between sinks and sources is not as consistent for the case of N$_2$-seeding. A nitrogen concentration of 10 - 25% could explain the mismatch between the ion target flux and the ionisation source prediction, assuming an average nitrogen ion charge of 2. A crude analysis, using Open-ADAS photon emission coefficients together with the NII (399.6 nm) line brightness measured by the DSS and ($T_e^E$, $n_e$, $\Delta L$) obtained from the Balmer line analysis, indicates the ratio between the N$^+$ density and n$_e$ is larger than 4%. The total nitrogen concentration is likely significantly higher than the N$^+$ concentration: to illustrate, for a transport-less plasma – which is not valid here – one would expect a fractional abundance of N$^+$ smaller than 0.1 for the values for $T_e^E$ obtained. This crude analysis is consistent with the explanation of a significant portion of the ion target current being due to nitrogen ions but does not constitute a proof. A proof would require a more quantitative and complicated analysis as in [63].

Volumetric recombination is found to be fully negligible during N$_2$ seeding (Figure 7f) [1]. This is consistent with results from other devices [8, 32] as well as MAST-U SOLPS simulations [64], where the role of volumetric recombination during N$_2$ seeded detachment has been found to be smaller.

### 3.3 Power balance in the divertor and relationship to ionization

We have now described all the elements in divertor particle balance and we will now investigate how this is related to divertor power balance. It has been suggested previously, both experimentally [8] and theoretically [11, 12, 17] that the ion source can be limited by the amount of power available for ionization in the divertor (which occurs simultaneously with momentum losses – section 4.2 & equation 2). To address directly whether power limitation of the ion source leads to the ion source behaviour we now develop a power balance analysis and apply it to the outer divertor for one of the discharges shown in Figure 7, #56567.

The power entering the divertor, P$_{div}$, is lost partially to radiation, P$_{rad}$, after which the remaining power ends up at the target (P$_{target}$), both in the form of potential energy, $P_{target}^{pot} = I_t \epsilon$ and kinetic energy, $p_{target}^{kin} = I_t \gamma T_t$, where $\gamma \sim 7$ is the sheath transmission factor. We use 'kinetic' to mean 'kinetic and thermal'. This is shown in Equation 6, where $\epsilon = 13.6$ eV is the potential energy. The molecular dissociation potential of 2.2 eV [3, 57] is neglected in dissociation, surface recombination and volumetric recombination.

$$P_{div} - P_{rad} = P_{target} = I_t(\gamma T_t + \epsilon) \tag{6}$$

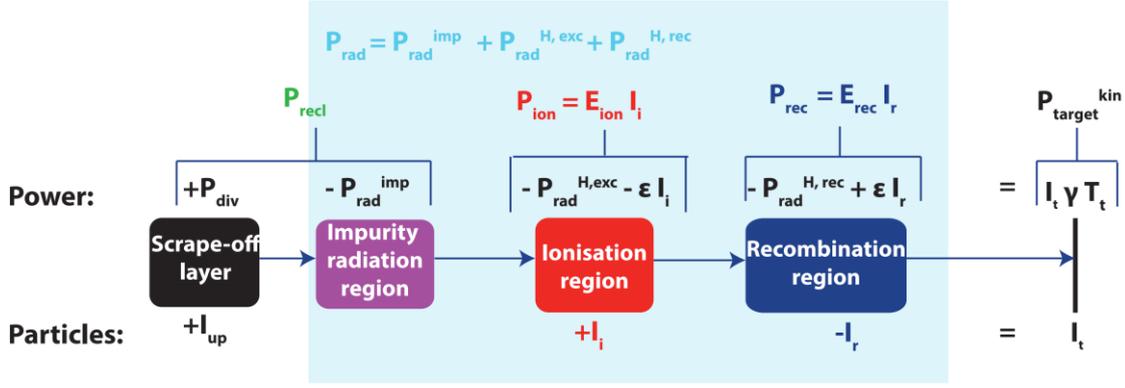

*Figure 8: Schematic overview of power and particle balance in the outer divertor. The blue shaded region represents the divertor. The top line describes power balance in terms of power entering the divertor ($P_{div}$), impurity radiation ($P_{rad}^{imp}$), power reaching the recycling/ionisation region ($P_{recl}$), ionisation power cost ($P_{ion}$) and recombination power cost ($P_{rec}$), eventually resulting in kinetic power reaching the target ($P_{target}^{kin}$). The bottom line in the Figure describes particles (ion) balance in terms of an ion flow from upstream ($I_{up}$), ions generated in the ionisation region ($I_i$) and ions being removed in the recombination region ($I_r$), eventually leading to $I_t$ ions reaching the target.*

The radiated power highlighted in Equation 6 can be split into different portions: hydrogenic radiation and impurity radiation ($P_{rad}^{imp}$) – equation 7. Hydrogenic radiation has both an excitation ($P_{rad}^{H,exc}$) and a recombination ($P_{rad}^{H,rec}$) contribution.

$$P_{div} - P_{rad}^{H,exc} - P_{rad}^{H,rec} - P_{rad}^{imp} = I_t(\gamma T_t + \epsilon) \qquad (7)$$

We ignore, for simplicity, energy losses due to charge exchange (CX) in equation 7. A large range of CX energy losses per ionisation event is suggested in literature (~ 3-5 eV [11]; 5-15 eV [1, 15, 65]). The CX power loss is not well known and is difficult to quantify with a simple model. Preliminary results from SOLPS simulations [35] indicate that CX related power losses are secondary to impurity and hydrogenic radiation and we do not include them in the following.

To obtain further insight into the power loss processes, we can re-arrange Equation 7 by bringing the potential energy of the ions reaching the target to the other side of the Equation and utilising the closed box approximation (Equation 5):

$$(P_{div} - P_{rad}^{imp}) - (P_{rad}^{H,exc} + \epsilon I_i) - (P_{rad}^{H,rec} - \epsilon I_r) = I_t \gamma T_t \qquad (8)$$

The grouped terms in equation 8 explain the different processes in divertor power balance and are schematically represented in Figure 8 as a visualisation aid. The different regions in Figure 8 spatially overlap as shown in section 3.1.2 (Figure 5) and the analysis does not rely on a separation of these regions. We will now discuss all these different grouped terms in detail.

The power flow into the divertor, $P_{div}$, is first reduced through divertor impurity radiation. We define which is left as $P_{recl}$: the power entering the recycling region (Equation 9).

$$P_{recl} = P_{div} - P_{rad}^{imp} \qquad (9)$$

Power is lost in the ionisation region by converting neutrals to ions (energy loss $\epsilon$ per ionisation) and by excitation collisions preceding ionisation leading to radiation losses ($P_{rad}^{H,exc}$): Equation 10. Dividing the total ionisation power loss $P_{ion}$ by the total ionisation source leads to an effective ionisation energy loss, $E_{ion}$ (Equation 11), which is an important parameter in modelling the ion target current dynamics as will be discussed in section 4.1 and 4.2. Furthermore, $E_{ion}$ will rise during detachment as the ionisation region grows colder and more excitation collisions occur before ionisation. This is observed experimentally (section 4.1.5) to have a significant influence on the ion current reduction during detachment (section 4.1.5) in agreement with 1D modelling predictions [66].

$$P_{ion} = P_{rad}^{H,exc} + \epsilon I_i \tag{10}$$

$$E_{ion} = \frac{P_{ion}}{I_i} = \frac{P_{rad}^{H,exc}}{I_i} + \epsilon \tag{11}$$

There are both energy gains (potential energy is released back to the plasma $\epsilon I_r$) and radiative losses ($P_{rad}^{H,rec}$) in the recombination process. The total power 'cost' of the recombination region ($P_{rec}$) is given by Equation 12, which is similar to how the power loss/gain due to recombination in [67] was determined. $P_{rec} < 0$ indicates a net plasma heating by volumetric recombination (e.g. $\epsilon I_r > P_{rad}^{H,rec}$).

$$P_{rec} = P_{rad}^{H,rec} - \epsilon I_r \tag{12}$$

The value of $P_{rec}$ on TCV is small, whether slightly positive or negative (see section 4.1.5, [1]). Recombinative plasma heating may occur in higher density/lower temperature devices where a significant portion of recombination is three-body recombination [4]. Throughout this work, 'effective recombination coefficients' are used which account for all types of recombination expected given the plasma conditions [53].

We now rewrite Equation 8 using the definitions of equations 9, 10, 12 into the newly derived terms:

$$P_{recl} - P_{ion} - P_{rec} = P_{target}^{kin} \tag{13}$$

As $P_{recl}$ is lowered through impurity radiation (while keeping $P_{div}$ constant and $P_{ion}$ rising), a point can be reached where $P_{recl}$ limits the power needed for ionization, $P_{ion}$. The ionization source, $I_i$, would then be reduced: 'power limitation'. A reduction of the ion source leads to less ions entering the recombination region (where more losses can occur) and thus a reduced target current, $I_t$. As part of this 'power limitation' process $P_{target}^{kin}$ also drops and the temperature near the target drops (see Figure 8 and Equation 8), making that region conducive first to ion-neutral collisions which are related to momentum loss processes, and then, as the target temperature continues to drop, recombination.

### 3.3.1 Power balance measurements

To utilize the divertor power balance structure described above we also need to explain how the various parameters are obtained experimentally. First, we start with determining the power flowing into the Scrape-Off Layer (SOL) from the core plasma, $P_{SOL}$. Since the discharges included in this study are Ohmically heated ($P_{Ohm}$), $P_{SOL}$ is obtained by subtracting the core radiated power ($P_{rad}^{core}$), measured by foil-bolometer arrays (see section 2 and [1]), from $P_{Ohm}$. The power flowing to the outer divertor is $P_{div} = \alpha P_{SOL}$, where $\alpha$ denotes the fraction of $P_{SOL}$ flowing to the outer divertor. In a previous study [68], it was found that the power asymmetry depends on poloidal flux expansion ($f_x$) and plasma current. Here $P_{in}/P_{out}$ (where $P_{in}$, $P_{out}$ is the power measured at the inner/outer strike points respectively) of ~1 was found for high plasma currents (340 kA, reversed field) and high values of poloidal flux expansion ($f_x \geq 6$) during attached and low divertor radiation conditions. Given that we assume for this investigation, $\alpha \sim 0.5$ for the plasma conditions (flux expansion; plasma current) of #56567 (340 kA, $f_x \sim 8$).

$$P_{div} = \alpha(P_{Ohm} - P_{rad}^{core}) \tag{14}$$

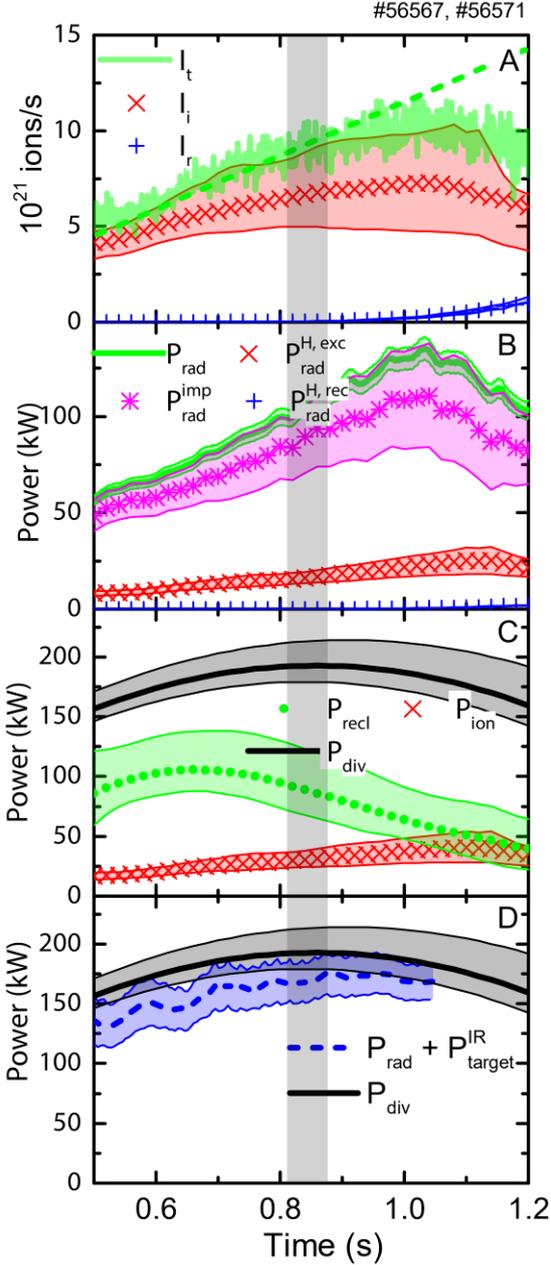

To obtain $P_{recl}$, first we need to estimate $P_{rad}^{imp}$, which requires separating out the impurity radiation from the hydrogenic radiation. Using Equation 15, this can be done by obtaining the divertor radiation, $P_{rad}$, from bolometry while estimating both the hydrogenic radiation $P_{rad}^{H,exc}$, $P_{rad}^{H,rec}$ through spectroscopic means [1, 2].

$$P_{rad}^{imp} = P_{rad} - P_{rad}^{H,exc} - P_{rad}^{H,rec} \qquad (15)$$

Figure 9 displays the result of our derivation of the various power channels for #56567; similar qualitative trends are found for the other two discharges presented in section 4.2. As the core density ramp proceeds, divertor radiation increases until the ion target current roll-over (Figure 9b). Impurity radiation is dominant (x 4) over $P_{rad}^{H,exc}$ with recombination radiative losses essentially ignorable. This impurity radiation results in a constant decrease of $P_{recl}$ during the core density ramp while $P_{div}$ remains roughly constant. This indicates (intrinsic, carbon) impurity radiation plays a key role, even in non-seeded density ramp discharges, in reducing the power reaching the recycling region in TCV, enabling detachment.

As the core density ramp proceeds, $P_{recl}$ and $P_{ion}$ grow closer together until $P_{recl} \sim 2 P_{ion}$ at the detachment onset (~0.83 – 0.87 s) and $P_{ion} \sim P_{recl}$ at the ion current roll-over (negative slope $I_t$ ~1.05 s) (Figure 9c). *This quantitative information suggests that the ion source is being limited by the power available, $P_{recl}$.* When $P_{recl}$ has dropped to roughly $P_{ion}$, $P_{target}^{kin} \ll P_{recl}$ – implying that low target temperatures are achieved as is expected from detachment and is observed (section 4.1.1). In that sense, $P_{recl} \sim 2 P_{ion}$ (e.g. $P_{target}^{kin} \sim ½ P_{recl}$) at the detachment onset is expected as some power, beyond ionization, is required to maintain a target temperature.

*Figure 9: Power balance investigation for the outer divertor for pulse #56567 a): ion target flux, ionisation rate and recombination rate; b) break-down of total radiation and its contributors; c) comparison between power entering outer divertor leg, $P_{div}$, the power entering the recycling region, $P_{rec}$, and the power needed for ionisation, $P_{ion}$; d) comparison between $P_{div}$ and the outer divertor leg radiative losses plus the measured power deposited on the target by the IR for consistency.*

Figure 9d includes a check of the overall divertor power balance. The sum of the total radiated power and the power reaching the target, $P_{rad} + P_{target}^{IR}$ (the latter term from IR measurements), is compared with the power flowing to the outer divertor region, $P_{div}$, and the two match within uncertainties, giving confidence in the $P_{div}$ determination. Note that $P_{rad} + P_{target}^{IR}$ is no longer shown after 1.05 s due to failures in the IR background subtraction algorithm.

## 4. Discussion

The results shown in section 3 of this paper show a strong particle balance correlation, in magnitude and time dependence, between the ionization source and the target ion current. This implies the ion current roll-over occurs due to an ion source reduction as opposed to an ion sink. In section 3.3 we also calculate that the power required to supply the measured ionisation source is approximately equal to the power flowing into the recycling region ($P_{recl}$); power limitation of the ionization source is occurring.

In the following discussion we utilize reduced analytic models to predict the detachment threshold and the accompanying target ion current (ion source) behaviour. These predictions are compared with observations and are used to investigate the relation between the target ion current and upstream parameters. Such reduced analytical models take the minimum number of necessary physical processes into account to model various detachment characteristics. In addition, the existence of other ion sources/sinks, apart from the ones treated in section 3, is also considered.

### 4.1 Investigating the ion target flux trends in the framework of power and particle balance

We now investigate the influence of 'power limitation' on the ion source more quantitatively by predicting the target ion source through its dependence on power and target temperature using power and particle balance [8, 12, 15] through the processes highlighted in Figure 8 & 9.

$$I_t = \frac{P_{recl} - E_{ion} I_r}{E_{ion} + \gamma T_t} = \left(\frac{P_{recl}}{E_{ion}} - I_r\right) \times f_{ion}(T_t^*) \tag{16}$$

$$f_{ion}(T_t^*) = \frac{1}{1+T_t^*} \tag{17}$$

The target ion current, $I_t$, is calculated using Equation 16, where $P_{recl}$, $E_{ion}$ and $T_t$ are the independent, measured, variables. Equation 16 is derived by combining the different power sinks presented in section 3.3 (Equations. 8, 11) with the closed box approximation (Equation 8). Though recombination is accounted for in particle balance, it is assumed that it is neither an energy sink nor an energy source (e.g. $P_{rec}$ ~ 0 in Equation 13), which agrees with spectroscopic estimates (section 4.1.5). The predicted $I_t$, in this form, applies to the entire (outer) divertor, although this model is also applicable along a single flux tube (neglecting cross-field transport of particles and heat). The target temperature, $T_t$, in Eqs. 19, 20 is therefore an effective averaged (weighted by the heat flux) target temperature [8], which is not necessarily representative of the peak temperature at the target.

$\frac{P_{recl}}{E_{ion}}$ represents the maximum ion source which could be achieved if all power entering the recycling region is spent on ionisation. In the absence of recombination[1], $f_{ion}(T_t^*)$ (Equation 17) represents the fraction of $P_{recl}$ spent on ionisation, in which $T_t^* = \frac{\gamma T_t}{E_{ion}}$ represents the ratio between kinetic power reaching the target and power used for ionisation ($\frac{\Gamma_t \gamma T_t}{\Gamma_t E_{ion}}$).

---

[1] Equation 16 can be re-written $I_t = \beta \times \frac{P_{recl}}{E_{ion}} \times \frac{1}{1+T_t^{**}}$ in which $\beta = \frac{I_t}{I_I}$ represents the fraction of ionised particles reaching the target and $\frac{1}{1+T_t^{**}}$ represents the fraction of $P_{recl}$ spent on ionisation even if recombination is present [35], in which $T_t^{**} = \beta \frac{\gamma T_t}{E_{ion}}$ represents the ratio between the kinetic power reaching the target and the power required for ionisation. For the case discussed, $\beta > 0.85$ and hence can be neglected.

It should be clearly noted that since Equation 16 requires $T_t$ as a measured *input*. It does not take *explicitly* into account that changing the power entering the recycling region also influences the target temperature.

### 4.1.1 Target temperature estimates

Obtaining the target temperature during detached conditions is challenging as $T_e$ measured by Langmuir probes often overestimated in detached low $T_e$ conditions [41, 59, 60].

An estimate of $T_t$ can be obtained spectroscopically from the line of sight closest to the target, which yields two different target temperatures [2]: one target temperature that is characteristic of the recombinative region ($T_t^R$) along the chordal integral and one target temperature that is characteristic of the excitation region ($T_t^E$) along the chordal integral. Those are both likely an upper limit with respect to the actual target temperatures as the chord views the separatrix region at ~5 cm above the target. As a consistency check, these spectroscopically-derived target temperatures are compared with a target temperature derived from power balance ($T_t^{PB}$ – Equation 18), which is obtained from Equation 7. Since $T_t^{PB}$ is obtained from the kinetic power reaching the target, $T_t^{PB}$ can be regarded as a heat flux averaged target temperature.

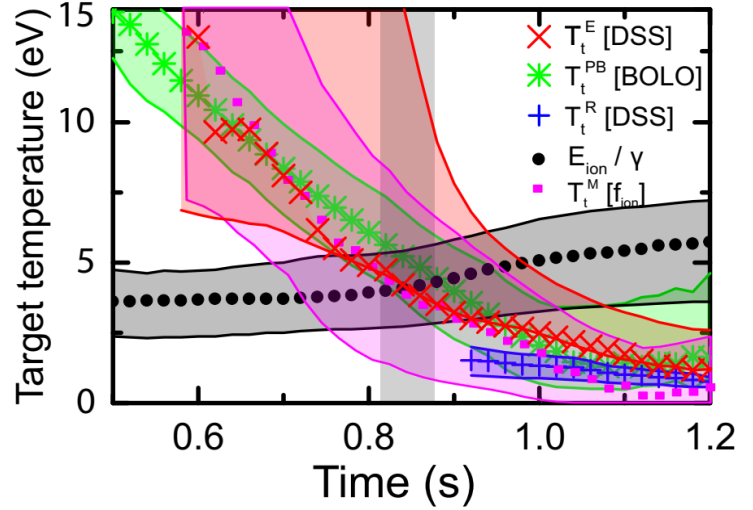

*Figure 10: Target temperature estimates from spectroscopy (excitation – $T_t^E$; recombination – $T_t^R$); power balance – $T_t^{PB}$ and power/particle balance modelling based on the measured $f_{ion} = P_{ion}/P_{recl}$ as function of time.; $T_t^m$ derived from $f_{ion}$. Discharge #56567.*

$$T_t^{PB} = \frac{P_{div} - P_{rad}}{\gamma I_t} - \frac{\epsilon}{\gamma} \qquad (18)$$

All three target temperature estimates show a decreasing trend as function of time, reaching target temperatures of 1-2 eV at the end of the discharge (Figure 10). $T_t^E$ and $T_t^{PB}$ agree within uncertainty, whereas $T_t^R$ (shown from 0.9 s onwards, since recombinative signatures are large enough to observe at this time) starts lower and decreases less strongly as function of time. $T_t^R$ is likely lower as recombination-dominated emission increases strongly at low temperatures and is thus dominated by contributions from lower-temperature parts of the plasma along the line of sight. We utilize $T_t^E$ in the following prediction of the target ion flux roll-over (Equation 16). This is appropriate as the excitation emission weighted temperature, $T_t$, is likely similar to the heat flux averaged temperature, as most excitation near the target occurs at the highest heat fluxes.

### 4.1.2 Comparing the measured and predicted ion target flux

Power and particle balances, as included in Equation 16, provide a quantitative prediction of the target ion current behaviour through the attached and detached periods for pulse #56567, discussed earlier in section 3.2 & 3.3. This requires four input parameters: First, $P_{recl}$ is derived from subtracting impurity radiation losses from the power entering the outer divertor (Equation 13, section 3.3) – using $P_{SOL}$, bolometry and hydrogenic radiation (obtained spectroscopically) estimates. The other three parameters are $E_{ion}$ (obtained using Equation 11, section 3.3), $T_t$ (for which we use $T_t^E$), and $I_r$ (section 3.1 - 3.2). All three latter parameters needed for equation 16 are directly determined through spectroscopic inferences [2]. The predicted ion target flux is in good agreement (in magnitude, trend and roll-over point) with experimental measurements of $I_t$ (Figure 11a, b). This shows that the ion target flux can be described fully in terms of the maximum possible ion source, $\frac{P_{recl}}{E_{ion}}$, and the recombination sink, $I_R$, once $T_t^*$ is known.

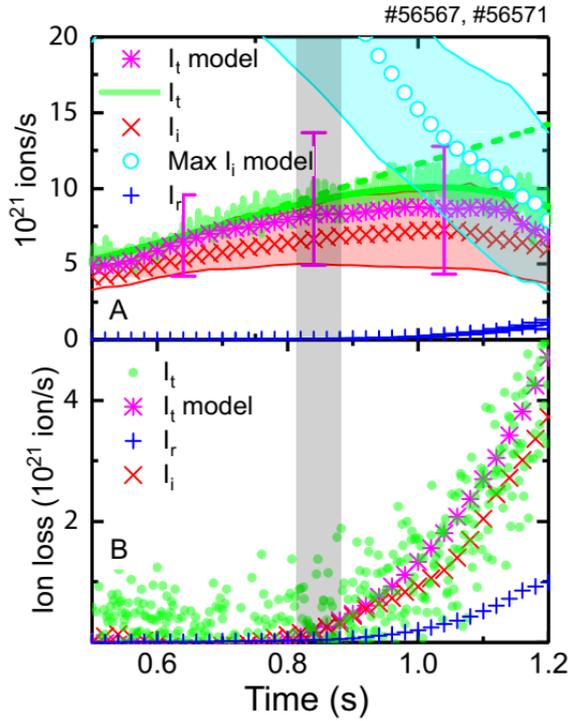

*Figure 11: a) Predicted ion target flux based on power balance compared with measured ion target flux as function of time. b) Ion loss (similarly defined as in section 3.1) as function of time for the ion target flux prediction, ion source and the measured ion target flux. Pulse #56567.*

The inferred maximum possible ion source, $\frac{P_{recl}}{E_{ion}}$, is of order 2x the measured ion source (Fig. 9), $I_i$, at the detachment onset (e.g. deviation of ion current trend from its linear reference which coincides with the roll-over of the separatrix ion target current for this particular discharge), which corresponds to $f_{ion} \sim 0.5$. This critical point is consistent with the empirical detachment threshold found in section 3.3, Figure 9 (0.83-0.87s). $f_{ion} \sim 0.5$ can also be written in terms of the target temperature (Equation 17): $T_t^* \sim 1 \rightarrow T_t \sim \frac{E_{ion}}{\gamma}$, which occurs when the black trend crosses the red trend in Figure 10; again consistent with the detachment onset.

The dynamics of the target ion current described by Equation 16 is a competition between two competing terms $-\left(\frac{P_{recl}}{E_{ion}} - I_r\right)$ and $f_{ion}$. $\left(\frac{P_{recl}}{E_{ion}} - I_r\right)$ decreases during a density ramp while $f_{ion}$ increases, driven by both the drop in $T_t$ and increase in $E_{ion}$ as the divertor cools. The increase in $f_{ion}$ is stronger in the period up to $f_{ion} \sim 0.5$, leading to a net increase in the target ion current before detachment. The increases in $f_{ion}$ are small after the detachment onset ($f_{ion} > 0.5$ & $T_t^* < 1$) and become insufficient to fully compensate the drop of $\frac{P_{recl}}{E_{ion}}$, resulting in a flattening of $I_t$. The target integrated ion current roll-over starts at a higher $f_{ion} \sim 0.7$, where $T_t \sim \frac{\epsilon}{\gamma} \sim 2$ eV. When $T_t$ reaches this level and drops further (e.g. $T_t^*$ approaches 0), one can approximate the target current as $I_t \sim (P_{recl}/E_{ion} - I_r)$. This observation is *operationally sufficient* to state that the ion source is becoming limited by the amount of power flowing into the recycling region; e.g. one can predict $I_t$ only given $P_{recl}/E_{ion}$ and $I_r$. In addition, when such temperatures are achieved, volumetric recombination can become a significant ion sink. All of this must, however, be consistent with a target pressure drop faster than $T_t^{1/2}$ (equation 2).

### 4.1.3 Comparing the measured and predicted power fractions of ionisation

The trends in $f_{ion}$ provide additional physical insight into the power dynamics of the recycling region and provides further means of comparing the power/particle balance model against experimental measurements. That is important as the functional form of $f_{ion}$ (equation 17), with the sheath target condition, leads to an analytic detachment onset [11, 12, 15] prediction at $f_{ion}$ = 0.5, as we will derive in section 4.2 and A.1.

First, as shown in Equation 17, $f_{ion}$ can be predicted based on $T_t^*$. $f_{ion}$ can also be inferred *directly* from the experimental spectroscopic observations and power balance as $f_{ion}$ = $P_{ion}$ /$P_{recl}$. The experimental inference (solid lines) agrees with the predicted $f_{ion}$ (symbols) within uncertainty (Figure 12b). Secondly, given a *measured* $f_{ion}$ and $E_{ion}$, one can model $T_t$ (labelled $T_t^m$) and compare it to the various experimental $T_t$ estimates. This comparison is shown in figure 10; indicating quantitative agreement within the uncertainty.

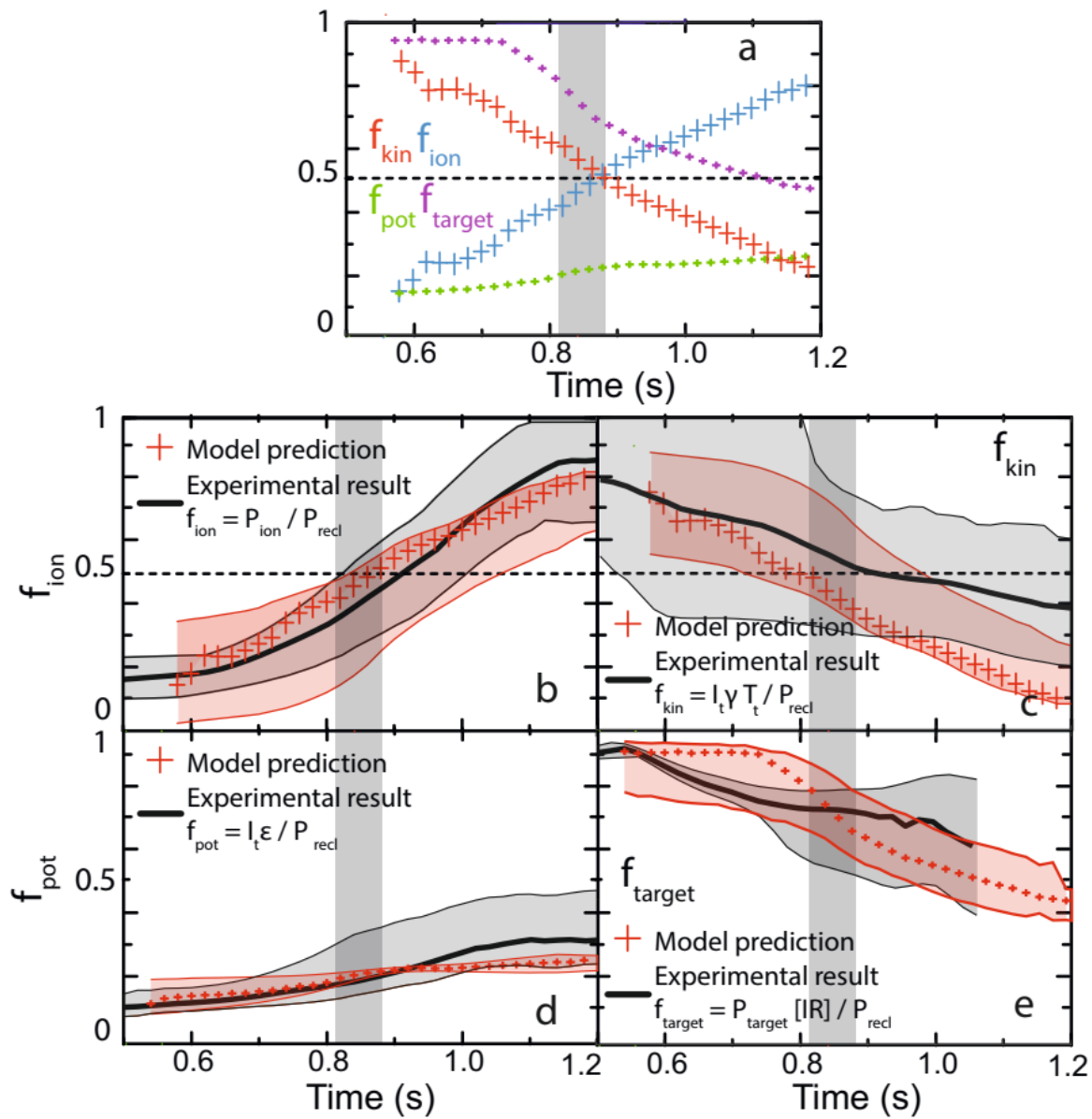

Figure 12: a) Break-down of the modelled fractions of $P_{recl}$ spent on ionisation; reaching the target; reaching the target in the form of potential energy and reaching the target in the form of kinetic energy (based on $T_t^*$; Eqs. 19, 21, 22). b-e) Comparison between the directly measured fractions with uncertainties and the modelled fractions (based on $T_t^*$; Eqs. 19, 21, 22) with uncertainties. Data obtained from IR imaging has been omitted from t>1.05 s due to a failure in the background subtraction algorithm.

Similar fractions to $f_{ion}$ can be derived, which model the fraction of $P_{recl}$ reaching the target in the form of kinetic/potential energy. These can be similarly compared to directly measured fractions to further validate the power/particle balance model, see Figure 12c-e. All of those directly measured fractions, analysed using a Monte Carlo probabilistic approach with uncertainties listed in [2], agree with the modelled power fractions as shown below indicating that a simple model based on $T_t$ and $E_{ion}$ can indeed predict the various power fractions. All modelled fractions are shown in figure 12a.

Since $f_{ion}$ is the fraction of $P_{recl}$ spent on ionisation, we can also calculate the fraction of $P_{recl}$ left after passing the ionisation region in the form of kinetic energy (Equation 19) – $f_{kin}$ which is compared with a 'directly measured' $f_{kin}$, which is $f_{kin}$ = $P_{target}^{kin}$ / $P_{recl}$ and $P_{target}^{kin} = I_t \gamma T_t$; $I_t$ is obtained from Langmuir probes and $T_t$ is the target excitation temperature estimated using spectroscopic analysis – Figure 12c.

$$f_{kin} = \frac{\gamma T_t}{\gamma T_t + E_{ion}} = 1 - f_{ion} = \frac{T_t^*}{T_t^* + 1} \quad (19)$$

The other fraction of $P_{recl}$ which reaches the target is power, in the form of *potential energy,* spent on neutral -> ion conversion (ionization) in the recycling region ( $\epsilon I_i$) assuming no volumetric recombination - provided by Equation 20. Again, this modelled $f_{pot}$ is compared with the directly estimated $f_{pot} = \frac{P_{target}^{pot}}{P_{recl}} = \frac{I_t \epsilon}{P_{recl}}$ in Figure 12c where $I_t$ measured by Langmuir probes was used.

$$f_{pot} = \frac{\epsilon/E_{ion}}{1 + T_t^*} \quad (20)$$

The fraction of $P_{recl}$ deposited at the target is the sum of the kinetic and potential terms: $f_{target} = f_{kin} + f_{pot}$, which decreases as function of time (Figure 12e) from 90% to 40%. That modelled value is compared with the measured $f_{target}$ = $P_{target}$ / $P_{recl}$, where IR measurements of the total power deposited at the target are used with an assumed 50% uncertainty on the measured $P_{target}$. When $f_{kin}$ approaches 0, $f_{pot}$ becomes the lower limit for $f_{target}$, and thus the power reaching the target, can attain: for $T_t^*$ -> 0 and $E_{ion}$ ~ 40 eV, $f_{target} \approx f_{pot}$ ~ 0.35. Volume recombination (or a further increase in $E_{ion}$) is required for a further reduction of $f_{target}$.

### 4.1.4 The case of power limitations and its implications

In the previous sections we have shown that an analytic model, accounting only *explicitly* for power/particle balance, using measurements of $P_{recl}$, $E_{ion}$, $T_t$ can predict several aspects of detachment in quantitative agreement with the experiment. We re-iterate that this must be consistent with target pressure loss (equation 2). Essentially, this is accounted for *intrinsically* as momentum losses play a role in the relation between the *measured* parameters $P_{recl}$, $E_{ion}$, $T_t$. Below we indicate we discuss the case of power limitation and its implications further.

One could imagine that the target ion current controls the upstream ion source as neutrals created at the target are needed for ionization upstream [11, 12, 15]. However, those neutrals would accumulate if $P_{recl}$ is not large enough to ionize them. This appears to be the case as the target ion current is strongly reduced in detachment, since the divertor neutral pressure (measured by baratron gauges [38]) stays high and even increases while the ion source is decreasing [38]. This is similar to C-Mod observations [69, 70]. SOLPS simulations [35] indicate that the neutral density averaged over the DSS chords (weighted by the excitation emission profile), as well as neutral pressures obtained in the simulation, increase during detachment while the neutral fraction ($n_0/n_e$) remains roughly constant [1, 2].

Some might suggest that the ionization source drop is not driven by limitation of the power available, but a natural consequence of low *target* temperatures (<5 eV) where the ionisation *probability* (e.g. the number of ionisations per neutral per electron per volume) is low – so fewer ions are created. However, such logic implicitly assumes that the neutral density and/or the power into that region is fixed. Another issue with that logic is that ionisation is a *volumetric* phenomenon and thus cannot be ascribed by only target parameters as the ionisation region expands from the target and can move all the way to the x-point.

It is true that as $T_t$ drops, $E_{ion}$ rises due to the additional excitations needed for each ionization and we point to this as a contributing factor in the loss in divertor ionization. However, using power/particle balance, we can make the statement that a low target temperature ($T_t < E_{ion}/\gamma$) *is a consequence of* power limitation of the ionisation source, as $P_{recl}/P_{ion} < 2$. Thus the direct cause of target ion current loss before recombination plays a significant role is power loss. This must occur coincidentally with target pressure loss (eq. 2), which is attributed (at least in part) to the formation of volumetric momentum losses – likely due to $T_t < 5$ eV.

### 4.1.5 The variation of $E_{ion}$ in detachment

Applications of analytic modelling often assume $E_{ion}$ to be constant [4, 8, 11]. The excitation radiation component ($\frac{P_{rad}^{H,exc}}{I_i}$) in $E_{ion}$ (Equation 11) is, however, strongly temperature dependant: as the divertor cools (<5 eV) more excitation occurs before ionization happens. During the density ramp, the effective temperature in the ionisation region drops, leading to a factor of two rise in $\frac{P_{rad}^{H,exc}}{I_i} = E_{ion} - \epsilon$, the radiation cost of ionization (see Equation 11 and Figure 13b), which results in a 50% increase in the divertor leg averaged $E_{ion}$ (weighted by the local ionisation rate).

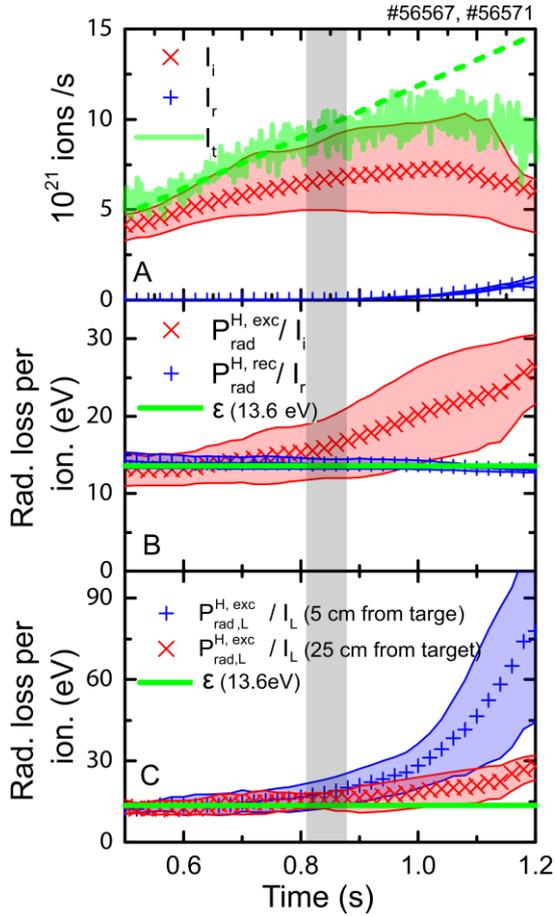

Both hydrogenic (through increasing $E_{ion}$) and impurity radiation (through lowering $P_{recl}$) can play significant roles in reducing the number of ionizations during detachment, despite the magnitude of hydrogenic radiation being much smaller than impurity radiation. This is demonstrated by the fact that the maximum ion source ($\frac{P_{recl}}{E_{ion}}$) decreases ~30% between t=1.0 and t=1.25 s, due to a ~10% decrease in $P_{recl}$ and a ~25% increase in $E_{ion}$. The importance of $E_{ion}$ was also raised in 1D modelling [71].

We also investigate the divertor profile of the excitation cost of ionization ($E_{ion} - \epsilon$) along different viewing chords, $\frac{P_{rad,L}^{H,exc}}{I_L}$. Poloidal temperature gradients lead to strong variations

*Figure 13: a) Target ion flux as function of time together with ionisation rate and recombination rate. b) Effective radiative energy cost per ionisation/recombination. c) Radiative energy cost per ionisation along a certain chord*

of $\frac{P_{rad,L}^{H,exc}}{I_L}$ along the divertor leg as shown in Figure 13c. In the region close to the target $\frac{P_{rad,L}^{H,exc}}{I_L}$ increases up to 80 eV. In hotter regions of the divertor leg (chords further away from the target), where most ionisation takes place (Figure 13b, c), the excitation radiation cost per ionization is 15-30 eV. Variations in geometry (e.g. closed vs open divertor, vertical- vs horizontal-target), which lead to variations in recycling and neutral penetration, could influence the location of the ionisation region and thus could affect the dynamics of the target ion current loss through a change of $E_{ion}$, amongst other changes.

Whether recombination can heat or cool the divertor plasma is determined by the competition between the energy loss due to recombinative radiation and the potential energy released back to the plasma upon recombination [9, 67]; as the plasma temperature and density vary the relative strength of two- and three-body recombination varies as well as the level of recombinative emission. For the TCV conditions investigated we find that the effective radiated energy loss per recombination event ($\frac{P_{rad}^{H,rec}}{I_R}$ Figure 13b) is roughly equal to the potential energy. That is not surprising, considering the modest TCV densities: ADAS calculations indicate an effective heating of 0 – 1 eV per recombination reaction at $T_e$ = 1 eV for $n_e$ in between $10^{18}$ - $10^{20}$ m$^{-3}$, using a similar calculation as done in [67] (e.g. $P_{rec}$ in Equation 12 divided by $R_L$). Hence, volumetric recombination does not lead to significant plasma heating or cooling for the TCV conditions presented.

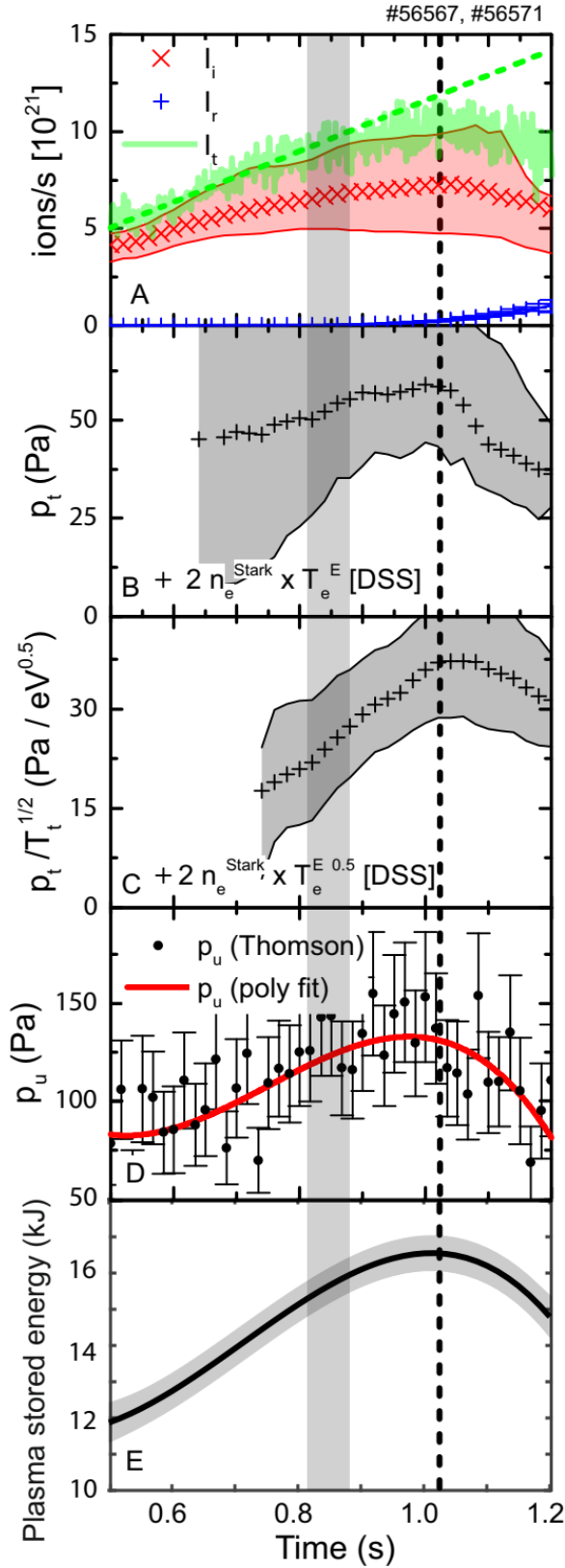

*Figure 14: Comparison of target pressure and upstream pressure. a) reference total current to the target, total ionization source and recombination sink; b-c) Target separatrix pressure ($p_t$) and ($p_t / T_t^{1/2}$) ratio based on spectroscopic measurements (Stark broadening + excitation temperature of the chord closes to the target); d) Upstream separatrix pressure from Thomson scattering; e) Plasma (core) stored energy from the diagmagnetic loop signal..*

## 4.2 Investigating the target ion flux trends in the framework of momentum balance

In the previous section we have investigated the target ion flux trend in the framework of power and particle balance of the entire SOL. In this section, we add momentum (pressure) balance [4, 9] to the power/particle balance analysis of section 4.1 such that the target temperature is now *predicted* instead of *set* by measurements. This enables a single flux tube comparison of the observed detachment dynamics and onset with additional predictions from simplified analytical theory; the preceding work has all been for the *entire* outer divertor. In this discussion, only the electron pressure is considered and the target pressure, $p_t$, is the total target pressure (e.g. twice the kinetic target pressure).

Trends in target ($p_t$) and upstream ($p_u$) electron pressure are compared in Figure 14. By assuming $p_t \sim p_u$ before detachment, $p_t$ appears to be significantly underestimated by ~ x2. Synthetic diagnostics through SOLPS [2] indicate that this difference is due to chordal-average nature of the spectroscopically estimated target pressure. Both the upstream and target pressure are observed to roll-over at the target ion flux roll-over (Figure 14b, d). The upstream density saturates simultaneous with a roll-over in the upstream pressure, while the upstream temperature drops (Figure 4). The reduction of the upstream pressure during detachment has also been observed in, at least one, other device(s): COMPASS [10].

It is striking that the roll-over of 1) the ion target current & divertor ionisation (figure 14a); 2) the upstream pressure (figure 14c) and 3) the plasma stored energy (figure 14e) all occur simultaneously within uncertainties. This may indicate that the cause of the upstream pressure roll-over during detachment is a deterioration of the plasma stored energy, which may be caused by enhanced core radiation.

As discussed in the Introduction (Equation 2), at any point during the discharge, the target ion flux scaling can be written as $\Gamma_t \propto p_t/T_t^{0.5}$, i.e. the target plasma pressure must drop faster than $T_t^{0.5}$ at the target ion flux roll-over. This is approximately observed (Figure 14c). However, $p_t/T_t^{0.5}$ is expected to deviate from linear from the detachment onset onwards, which is not the case. That discrepancy could be due to line integration effects as explained previously. $\Gamma_t \propto p_t/T_t^{0.5}$ (equation 2) links the trend in the ionisation source (section 4.1 - Equation 16) to the trend in the target pressure (Equation 2) and is thus crucial for understanding the complex interplay between momentum balance and ionisation balance.

### 4.2.1 Modelling total target ion current behaviour with both power and momentum balance

We utilise a 'two point' divertor model [4], which accounts for hydrogen recycling energy losses, to model the total target ion current. We refer to this as the '2PMR', discussed previously in literature [4, 11] and more extensively in [9]. See appendix A.1 for a full derivation and in A.2 we demonstrate how we apply and evaluate the 2PMR. Our first goal of the application of the 2PMR is to verify the *expected* ion target flux trend in *attached* conditions. For this, pressure constancy along a flux tube is assumed and since $p_u$ is a set *input* to the 2PMR, the target pressure $p_t$ is fixed as well. Under such conditions the 2PMR provides two possible solutions: one stable and one unstable. We assume in the following (section 4.2.1, 4.2.2) that the unstable solution *cannot* occur.

The 2PMR yields a relation for the target temperature (Appendix A.1 Equation A.5) as a function of $E_{ion}$ and $\frac{p_u}{q_{recl}}$ [4, 9, 11]. We obtain the flux tube specific $q_{recl}$ from $P_{recl}$ (which is for the entire outer divertor) by assuming its shape is exponential with the same SOL width as the measured IR heat flux profile (see appendix A.2). The 2PMR-predicted $T_t$ can be used to predict the target ion flux density ($\Gamma_t$ in ion/s m²) on a *single* flux tube, as shown in Equation 21 (Appendix A.1 Equation A.8). It is important to note that the 2PMR $\Gamma_t$ relation (Equation 21a) is *identical* to the flux surface equivalent of Equation 16 (while using the 2PMR predicted $T_t$ - Appendix A.1), which is shown in Equation 21b for reference to the reader. The 2PMR thus connects the standard two-point divertor model with the power/particle balance model discussed in section 4.1. Here $f_{ion}\left(T_t(E_{ion}, \frac{p_u}{q_{recl}}), E_{ion}\right)$ denotes that $f_{ion}$ (or $f_{kin}$) is a function of $E_{ion}$ and $T_t$, which is a function of $E_{ion}$ and $p_u/q_{recl}$ (assuming pressure balance).

$$\Gamma_t = \frac{\gamma p_u^2}{2\,m_i q_{recl} f_{kin}\left(T_t(E_{ion}, \frac{p_u}{q_{recl}}), E_{ion}\right)} \quad (21a)$$

$$\Gamma_t = \frac{q_{recl}}{E_{ion}} \times f_{ion}\left(T_t(E_{ion}, \frac{p_u}{q_{recl}}), E_{ion}\right) \quad (21b)$$

To compare the experimental measurement of the *total* target ion current $I_t$ (as opposed to $\Gamma_t$) to the 2PMR, we integrate $\Gamma_t$ across the SOL. $I_t$ can then be modelled using $\frac{p_u^2}{P_{recl}}$, $f_{kin}$ ($T_t$) and $f_p$. $f_p$ parametrises the influence of the $p_u$, $q_{recl}$ spatial profiles as well as the divertor magnetic geometry on $I_t$. More information is provided in appendix A.2.

$$I_t = \frac{2\gamma\pi^2}{m_i} \times \frac{1}{f_{kin}} \times f_p \times \frac{p_u^{0\,2}}{P_{recl}} \quad (22)$$

The observation that $I_t$ scales linearly with the upstream density in contrast to the DoD scaling has been pointed out previously [38]. Both the trend and absolute value of the ion target flux prediction by Equation 22 agrees with the measured target flux in the attached phase as shown in Figure 15a, showing a clear linear increase as function of time (and thus upstream density – section 3.1.1). Hence, simply using $I_t \propto n_{e,u}^2$, on which the "Degree of Detachment" (DoD) [7], a parameter often used to investigate the magnitude of the ion current loss during detachment [24, 38] is based, is not appropriate for the TCV density ramp discharges studied. Similar in-depth investigations, where changes in $p_u$, $q_{recl}$, $E_{ion}$ (or even $p_u$, $q_t$) have been accounted to model the expected $I_t$ trend have not (yet) been carried out on other devices. It is thus unknown whether the often assumed $I_t \propto n_{e,u}^2$ is valid for other machines; the scaling $I_t \propto n_{e,u}^2$ should be verified through the more complete analytic model of equation 22 before the simplified DoD based on assuming $I_t \propto n_{e,u}^2$ is utilised.

In Equation 22, the main influence on $I_t$ is $\frac{p_u^2}{P_{recl}}$ (see Figure 15b for all the terms). This basic scaling (or $\Gamma_t \propto \frac{p_u^2}{q_{recl}}$) not only arises from the 2PMR, but can also be obtained from pressure balance ($n_u T_u = 2 n_t T_t$), the sheath target Equation ($q_t \propto n_t T_t^{3/2}$) and an equation for the (kinetic) target heat flux ($q_t = \Gamma_t T_t$). This results in $\Gamma_t \propto \frac{n_u^2 T_u^2}{q_t}$ (equivalent to Equation 5.13 in [4] of the basic two point model), providing an identical relation for the target ion flux as Equation 21a. Additionally, the $\Gamma_t \propto \frac{n_u^2 T_u^2}{q_t}$ scaling is also equivalent to the relation used in [7] for defining the degree of detachment originally (which is obtained by using equations 3,4,8 in [7]). Considering that scaling, even if $T_u$, $q_{recl}$, $E_{ion}$ are held constant and only the upstream density is increased, a different scaling than $\Gamma_t \propto n_{e,u}^2$ is expected when recycling energy losses are accounted for as the power flux required for ionisation is increased at higher densities, reducing $q_t = q_{recl} f_{kin}$ in the process.

Since $I_t \propto \frac{p_u^2}{P_{recl}}$ increases linearly as $n_u$, $\frac{T_u^2}{P_{recl}}$ must decrease roughly as $1/n_u$. Given that $P_{recl}$ decreases during the density ramp (Figure 9), $T_u$ (Figure 10a) must decrease more strongly to give this scaling. As $P_{div}$ is roughly constant throughout the discharge (Figure 9) a decrease of $T_u$ could result from an increase in cross-field energy transport in the SOL (SOL broadening is measured by IR thermography to increase [44] by over a factor 3 until detachment is reached for the discharge shown). Alternatively, the decrease of $T_u$ could be partially due to an increase in convective over conduction parallel heat transport [4]. Such trends could be different in TCV to other devices, which is subject to further investigation. This decrease of $T_u$ is qualitatively consistent with TCV SOLPS modelling [35, 56], which reproduces the linear $I_t$ (or scaling $\Gamma_t$) in attached conditions. It is, however, likely not related to the open geometry of the TCV divertor as this reduction of $T_u$ during a density ramp is also observed in TCV baffled SOLPS modelling [56]. MAST-U SOLPS-ITER simulations also indicate a $I_t \propto n_{e,u}$ trend in the attached phase [64, 72].

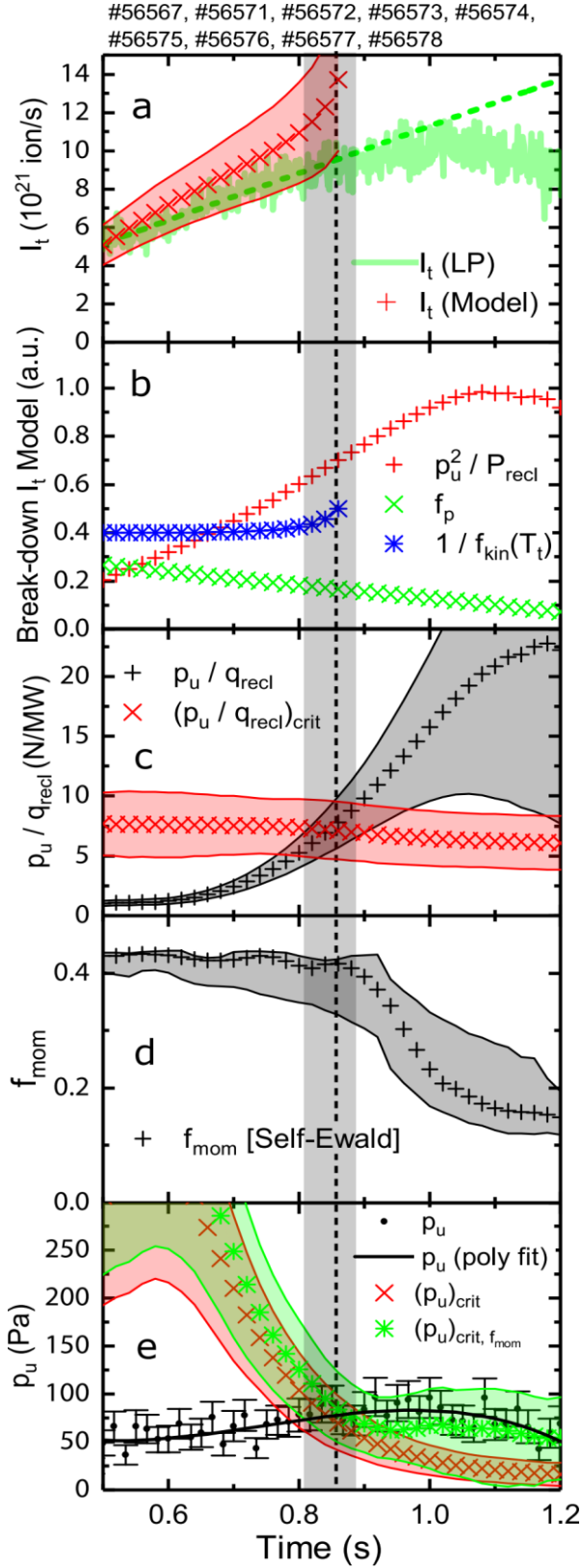

Figure 15: a) Measured and predicted target ion flux trend. b) Break-down of the contributors to the predicted ion flux. c) Measured $p_{up}/q_{recycl}$ compared to the critical predicted level. d) Inferred momentum losses from spectroscopic estimates. e) Measured upstream pressure compared to critical pressure level with and without momentum losses.

A previous TCV study concluded that the observed linear trend of $I_t$ with $\overline{n_e}$ indicated that the divertor plasma was in a low-recycling operation [37]. However, given our measurements that the target ion current is dominated by the divertor ionisation source (section 4.2) and that $I_t$ is properly predicted by the 2PMR (which assumes that all target ion current is due to divertor ionisation) strongly supports a characterization of the divertor as high-recycling.

### 4.2.2 Detachment thresholds and implications for momentum/pressure losses along a flux tube (separatrix)

As shown in Figure 15a, the 2PMR (where, for this case, we assume both constant pressure along the field line and a prescribed upstream pressure) can only be used to estimate $I_t$ until 0.8 s, at which time the 2PMR, under these assumptions, no longer obtains a solution (appendix A.1).

It is evident from Equation 2 that the ion current roll-over, together with a fixed/decreasing target temperature, *must* be accompanied by *target pressure loss* ($p_t \propto \Gamma_t T_t^{1/2}$). The power and particle balance model discussed in section 4.1 (eqs 19,20), indicates that *$\Gamma_t$ is a function of $T_t$* as there is a trade-off between using power for ionisation and the power flowing to the target expressed by $f_{ion}$ – Equation 17. Combining this with $p_t \propto \Gamma_t T_t^{1/2}$ has two implications: 1) the target pressure cannot be increased indefinitely, and a maximum exists ($p_{t,crit} = \frac{q_{recl}}{2\gamma c_s(T_t=\frac{E_{ion}}{\gamma})}$), 2) this maximum is reached at a certain threshold ($T_t = \frac{E_{ion}}{\gamma}$) where further decreases in $T_t$ lead to a smaller increase in $\Gamma_t$ than predicted by the decrease in $1/T_t^{1/2}$. That critical maximum target pressure (or threshold) is reached at ~0.83-0.87 s. The changed relationship between $\Gamma_t$ and $T_t^{-1/2}$ must be provided by a drop in $p_t$; this corresponds to the deviation of the ion current from its linear (attached) trend. Solutions beyond this point are not allowed by the model assumptions of a fixed $p_u$ and constant pressure along the field lines. It is thus not surprising that the 2PMR, under these assumptions, cannot model the ion current roll-over and thus only applies to *attached* conditions.

The above threshold of the 2PMR model is where the target pressure is maximised and target pressure loss is necessary as $\Gamma_t$ ($T_t$) starts to rise slower than $T_t^{1/2}$ (Equation 16, 17), has been suggested by Krasheninnikov [11, 15] to be a 'detachment onset criterium': for target temperatures below this limit insufficient power is transferred beyond the ionisation region to sustain a sufficiently high target temperature for the target pressure (which is collapsing) to match the upstream pressure. Stangeby, although not calling the above limits a detachment threshold, argues properly that to reach $T_t < E_{ion}/\gamma$, processes which continuously lower the target pressure as the target temperature becomes lower (e.g. the target pressure must be a specific function of the target temperature) is required [4, 9]. This can be achieved by volumetric momentum losses (as shown explicitly in [9]) and/or by assuming $p_u$ drops as function of $T_t$. See appendix A.1.3 and [1] for more information.

These 'detachment thresholds' can be written in three different forms, given by Equation 23, as shown in Appendix A.1. Without any knowledge on momentum loss, these thresholds represent the *lowest temperature* at which *target pressure loss must occur.* We expect these thresholds to correspond to the detachment onset as target pressure loss must occur at the detachment onset.

We have found thresholds given in equations 23b,c experimentally (section 3.3, 4.1, Figures 9,11) to be empirical thresholds for detachment. A third (equivalent) threshold (Equation 23a) can be derived from the 2PMR (Appendix A.1) [11] providing a critical maximum target pressure $p_{t,crit} = \frac{q_{recl}}{2\gamma c_s(T_t=\frac{E_{ion}}{\gamma})}$. Under the assumption of pressure balance, this is commonly written [4, 11] as a critical threshold for $p_u / q_{recl}$, above which $p_u / q_{recl}$ cannot rise (Equation 21a - assuming $p_t = p_u$), where $c_s$ is the target sound speed at $T_t = \frac{E_{ion}}{\gamma}$. ($p_t / q_{recl}$)$_{crit}$, which applies to a flux tube – not the average over the divertor, is compared to the experimentally inferred $p_u/q_{recl}$ in Figure 15c. The increase in $p_u/q_{recl}$ is mostly ascribable to a drop in $q_{recl}$ during the pulse.

$$p_{t,crit} = \frac{q_{recl}}{2\gamma c_s\left(T_t=\frac{E_{ion}}{\gamma}\right)} \rightarrow \left(\frac{p_t}{q_{recl}}\right)_{crit} = \frac{1}{\gamma c_s(T_t=\frac{E_{ion}}{\gamma})} \quad (23a)$$

$$T_{t,crit} = \frac{E_{ion}}{\gamma} \quad (23b)$$

$$f_{ion} = \frac{1}{2} \quad (23c)$$

This third critical limit (eq. 23a), evaluated at the separatrix, is also reached at the detachment onset (~ 0.83-0.87s, where the integrated target ion current starts to flatten and deviates from the linear trend) (Figure 15c). This is similar to the other detachment criteria ($f_{ion}$= 0.5 (Figure 12a), $T_t = \frac{E_{ion}}{\gamma}$ (Figure 11)). As discussed earlier, 0.83-0.87s also corresponds to where the separatrix current *density* starts to roll-over for this particular case (Figure 4b).

### 4.2.3 The 2PMR and pressure losses

The measured $p_u/q_{recl}$ rises above the ($p_t/q_{recl}$)$_{crit}$ threshold (figure 15 c), which is indicative of volumetric momentum losses causing a separation between $p_u$ and $p_t$. Defining momentum loss by $p_u f_{mom} \equiv p_t$, then the separation of ($p_u$)$_{crit}$ and ($p_t$)$_{crit}$ is accounted for by $1/f_{mom}$ (Equation 24). This equation implies that, given a known amount of $f_{mom}$ and $p_{t,\,crit}$, there is a certain *maximum upstream pressure limit* consistent with those two parameters.

$$p_{u_{crit}} = \frac{p_{t_{crit}}}{f_{mom}} = \frac{1}{q_{recl} f_{mom} \gamma c_s\left(T_t = \frac{E_{ion}}{\gamma}\right)} \quad (24)$$

From Equation 24 we find that $f_{mom}$ would need to start to decrease from 1 at the detachment criterion to ~0.4 at the end of the discharge to match the measured $p_u/q_{recl}$ (Figure 15c) to $(p_u/q_{recl})_{crit}$ in the detached phase using Equation 24. Such momentum losses in the TCV divertor during similar experiments have been determined directly from upstream and target pressure measurements [73], implying momentum losses greater than 50%.

An independent estimate of the onset and magnitude of momentum losses based solely on the dominance of charge exchange over ionisation can be made using the well-documented Self-Ewald model [5, 74] (Equation 25) which has been used in several other studies. Such an estimate assumes that the charge exchange rate equals the momentum loss rate; *e.g. each CX reaction leads to a complete loss of that ion's momentum,* which is an overestimate. Although the Self-Ewald model is an oversimplified momentum loss model, it does yield results in fair agreement with experiments and simulations [5, 15, 74].

That agreement may arise 'accidentally' from the temperature trend of $f_{mom}$ predicted through the Self-Ewald model, rather than the Self-Ewald model predicting the underlying physics correctly. The Self-Ewald model does not account for other momentum sinks, such as molecular-ion collisions [74, 75] which could supply the over-estimated CX momentum losses. Although the level of momentum loss due to molecules is unknown for TCV, we do know that molecules are present and undergoing reactions in the volume of TCV from simulations [35] as well as experimental measurements of $D\alpha$ [1]. Momentum loss can also occur due to recombination. However, from a simple SOL model [76] we have evaluated the reduction of $f_{mom}$ due to recombination for the case studied and found it negligible (smaller than 1.5% – in agreement with results from [8]). In addition, differences in transport could contribute to the observed and simulated pressure loss during detachment – or instance, cross-field transport may 'smear-out' pressure across the field lines, leading to a reduction in the high pressure regions near the separatrix [74].

With those caveats in mind, we integrate the spectroscopically–determined profile of charge-exchange and ionisation rates along the outer divertor leg (Figure 5e) to calculate the Self-Ewald $f_{mom}$ as a function of time, Figure 15c. We thus derive $f_{mom}$ from measurements, as opposed to a prescribed function $f_{mom}(T_t)$ as used in [4, 9, 77], which is unknown for TCV. This approach supports a roll-over in the 2PMR but does not support $T_t < \frac{E_{ion}}{\gamma}$, which would require a $f_{mom}(T_t)$ parametrisation (see section A.1 and [1] for more information).

As shown in Figure 5e, charge exchange to ionisation ratios are higher near the target during detachment than elsewhere in the divertor which, in the Self-Ewald model, results in larger inferred momentum losses. Note that we use the local temperature (excitation), charge exchange and ionisation rate estimates obtained spectroscopically for each chord, instead of the target temperature (used in other studies) which we feel more accurately represents what is occurring; using the target temperature would have led to larger inferred momentum losses. Furthermore, SOLPS simulations for TCV indicate that volumetric pressure loss can occur in the volume of the divertor [35]; not just in front of the target as observed in simulations [77] for other machines, which may invalidate making $f_{mom}$ a function of the target temperature.

$$f_{mom} = 2\left(\frac{\alpha}{\alpha+1}\right)^{\frac{\alpha+1}{2}} \quad (25)$$

$$\alpha = \frac{1}{1 + \frac{\int 2\pi r_i \, CX_L(r_i, z_i) dz}{\int 2\pi r_i \, I_L(r_i, z_i) \, dz}}$$

Our estimate of $f_{mom}$, using the Self-Ewald model, drops from ~0.9 to ~0.3 ([0.2 – 0.4] with uncertainty) is shown in Figure 15c, in agreement with the momentum losses obtained experimentally [62, 73] and with the $f_{mom}$ required to explain the increase of $p_u/q_{recl}$ beyond the $p_t/q_{recl}$ limit discussed above. This may be a coincidence – as mentioned above, the reality of CX collisions not carrying away 100% of the ion momentum may be compensated by ion-molecule collisions (not included) carrying that momentum away [75, 77].

### 4.2.4 The case for divertor processes reducing the upstream pressure and density

The results of the previous section show that the rise of $p_u/q_{recl}$ beyond its critical $p_t/q_{recl}$ limit can be attributed, at least partially, to momentum losses. However, $p_u$ also drops during the detached phase. The question of what leads to the drop in upstream pressure (and density) during detachment has been discussed by several authors of analytic and modelling studies [11, 12, 14]. In section 4.2 (before section 4.2.1) we have already shown that this reduction in upstream pressure is correlated with a reduction of stored energy in the core. This is one of the several possible explanations that may explain the reduction of the upstream pressure. The reduction of the upstream pressure and the role it plays in matching $p_u/q_{recl}$ to its critical threshold (Equation 24) is discussed here further.

Recombination has been predicted to lead to saturation of the upstream density when its rate approaches the ionization rate in a flux tube through a feedback loop [78]: as $n_u$ increases, the divertor cools further, hence augmenting the recombination sink and moving the recombination region further towards the x-point, potentially impeding a rise in $n_u$ [78]. This is not the case for these TCV discharges as recombination remains low and can be negligible. In addition, the recombination region peak does not move far off the target (Figure 5, [31]).

Krasheninnikov [11] offers another explanation for saturation of the upstream density. During detachment, insufficient momentum losses along flux tubes can constrain, or pull down the upstream pressure. It is important to reiterate that, although an $I_t$ roll-over requires a *target pressure* drop which increases faster than $T_t^{1/2}$ (Equation 2), analytically (from the viewpoint of the 2PMR) this can be provided by *either* volumetric momentum loss *and/or* a reduction of upstream pressure (Appendix 4). However, experimentally, one would like to avoid a degradation of the upstream pressure in reactor-relevant divertor solutions as this can influence the core plasma [4]. This requires $p_u \gg p_t$ and necessitates volumetric momentum loss to reduce $p_t$.

Using Equation 24, we make a direct comparison between the measured (Thomson scattering) $p_u$, the maximum upstream pressure limit $p_{u,crit}$ and $p_{t,crit}$ (which, in the case of no momentum loss – $f_{mom} = 1$, equals $p_{u,crit}$) as a function of time (Figure 15e). The measured upstream pressure rises during the density ramp, while $p_{u,crit,fmom}$ and $p_{t,crit}$ drop due to a decrease in $q_{recl}$ until $p_u$ crosses $p_{t,crit}$ and $p_{u,crit}$ at ~0.8 s, the detachment onset point. After that time the target pressure limit decreases further while volumetric momentum losses start to result in a bifurcation between the upstream/target critical pressures. Despite this bifurcation, $p_{u,crit}$ flattens and eventually rolls over, while $p_u$ continues to track $p_{u,crit}$. This indicates that, even when considering the amount of observed momentum loss, the observed saturation/roll-over of the upstream pressure is *consistent* with the model.

Detachment requires target pressure loss (eq. 2) which could be engendered by volumetric momentum loss and/or upstream pressure loss. Our experimental measurements and analysis using the 2PMR imply, given the amount of volumetric momentum loss, *a saturation/reduction of the* *upstream pressure* is required as well to reach $p_{t,crit}$. This is consistent with the *measured* reduction of the upstream pressure. However, this *consistency* does not indicate *causation:* e.g. it does not show that inadequate momentum loss on a given flux tube *causes* the upstream pressure to drop. As suggested earlier, other processes, i.e. upstream or divertor cross-field transport (particles and/or momentum), may be affecting the upstream pressure as well.

A commonly held assumption is that the upstream pressure remains constant/unaffected by detachment. That assumption results in the (mis)understanding that *all* the required $p_t$ drop *must* be provided by *only* volumetric momentum losses. These TCV results, however, show that *both* an upstream pressure drop and volumetric momentum losses contribute to the required $p_t$ drop. Accounting for upstream pressure changes is thus crucial for understanding detachment.

### 4.2.5 The role of momentum loss and upstream pressure loss in target ion current loss

As described in the introduction, researchers generally look at detachment with different emphases: power/particle balance and momentum balance, which mostly focusses on volumetric momentum losses. As explained earlier, the 2PMR, combines both points of view and in the 2PMR both approaches are equivalent (equations 21a & b, derives in A.1 equation A.8). The 2PMR predicts detachment occurs when power limitation starts ($P_{ion}$ ~ ½ $P_{recl}$; $T_t$ ~ $E_{ion}/\gamma$ ~ 4-6 eV), which corresponds to the point where the ion target current increases slower than $1/T_t^{0.5}$, hence *requiring* a target pressure loss. Thus, both target pressure loss and power limitation are required for detachment in 'high recycling' condition. For a demonstration of the equivalence of pressure and power balance points of view we refer the reader to Equation 21a & b, which was derived in the appendix as Equation A.8. which shows the 2PMR can be seen from either a power/particle or pressure balance description – which are equivalent in this model.

It is striking that the temperatures ($T_t < E_{ion}/\gamma$ ~ 4-6 eV) at which target pressure loss *must* occur (2PMR), according to divertor-physics, corresponds to the temperatures at which volumetric momentum loss *can* occur, according to atomic physics. This seeming coincidence of divertor and atomic physics implies volumetric momentum loss develops when power 'limitation' conditions ($P_{recl} < 2P_{ion}$) are reached, implying that power 'limitation' is a requirement for detachment for both points of view discussed.

The results of Section 4.2.4 show that the commonly held assumption that the upstream pressure remains constant/unaffected by detachment is not always true. Instead, the upstream pressure, target pressure and any volumetric momentum loss must be consistent with each other. This means the role of volumetric momentum loss can only be fully understood if all the processes influencing the upstream pressure are understood. These may be divertor, scrape-off layer and core processes. Examples could include changes in cross-field transport of energy, momentum and particles or volumetric losses within a flux tube, or both. The reality, however, is that we lack a quantitative understanding of how $p_u$ is influenced by both the core and divertor plasma, which likely requires an integrated core-edge model. Lacking such a model prevents us from fully ascertaining the role momentum loss plays in detachment.

Nevertheless, it is unlikely that momentum losses directly reduce the ion target current during fully power-limited ($P_{recl}$ ~ $P_{ion}$) detachment as $I_t$ ~ $P_{recl}/E_{ion}$ (section 4.1) for those conditions: momentum losses slow down the fluid velocity in a flux tube, but do not directly reduce the ion flux through the tube. Target pressure loss is, however, a necessity to reach fully power-limited conditions. Momentum losses may, in addition, facilitate detachment indirectly by allowing higher upstream pressures, leading to higher divertor densities (for the same $T_e$) and thus higher divertor radiation and higher recombination rates.

## 4.3 Investigating additional ion sources and sinks

Although the ionisation source and the volumetric recombination sink within the divertor are sufficient to explain, within uncertainties, the target ion flux trend (section 3.2), additional ion sources and sinks may remain significant.

In this work we have assumed that ion sources outside the divertor, leading to ion flows into the divertor ($I_{up}$), can be neglected as a source of ions reaching the target. While direct measurements of ion flows into the divertor are unavailable, we can estimate such flows through Equation 26, in which $M_u$ is the upstream Mach number and r the radial distance from the separatrix [79], which includes fluid flows along the magnetic field but ignores several types of drift flows such as ExB flows. Furthermore, this investigation ignores the influence of ionisation in the scrape-off-layer explicitly.

$$I_{up} \approx 2\pi M_u \frac{B_p}{B_t} \int_{seperatrix}^{wall} r\, n_u(r) \sqrt{\frac{T_u(r)}{m_i}} dr \qquad (26)$$

To estimate the *maximum* possible $I_{up}$, we use $M_u \sim 0.5$, the upper bound of a previous survey of upstream Mach number profiles across three tokamaks [79]. To compute this conservatively large $I_{up}$ (Equation 26), separatrix upstream densities and temperatures were measured using Thomson scattering, while their profiles were measured with a reciprocating probe (details are provided in appendix A.2). The resulting $I_{up}$, shown in Figure 16 for the high current density ramp discharge previously discussed in sections 3 and 4.1, 4.2, increases during the core density ramp which, of course, also raise the SOL density and thus $I_{up}$. $I_{up}$ remains small compared to the divertor source of ions and the target ion current except at the end pulse when recombination starts to become significant and the target current has rolled over. During detachment, $I_{up}$ increases to ~30% of the ion (outer) target flux and ion source.

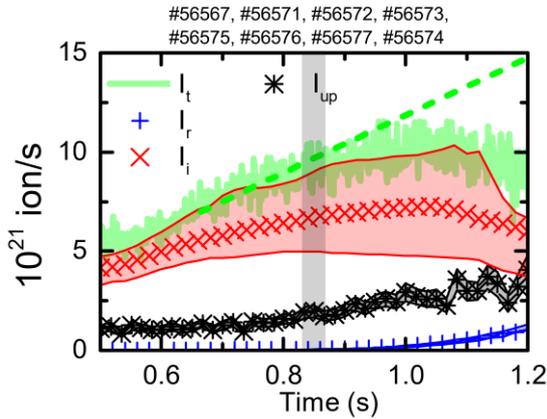

*Figure 16: The target ion flux compared to the ion source, inferred ion flow from the SOL and the recombination rate.*

The overall particle balance (Equation 27) would be consistent with the addition of our estimated $I_{up}$ within uncertainties. Even with $I_{up}$ included, the ion source, $I_i$, remains the largest contributor to the target ion flux and its roll-over after the detachment onset. Based on the measurements shown in section 3.1; it is likely that this is the case also for lower current (e.g. lower density) discharges.

$$I_t = I_{up} + I_i - I_r \qquad (27)$$

Ionisation in the scrape-off-layer increases as the core density is increased according to SOLPS simulations [18, 35, 56], contributing to $I_{up}$. This ionisation occurs either from recycled neutrals from the main chamber or from escaped neutrals from the divertor.

Other possible sources/sinks that could affect eq. 27 are molecular activated ionisation sources (MAI) and sinks (MAR). Evidence for molecular reactions which may lead to MAI/MAR has been found from the measured $D\alpha$ in TCV [1]. The measurements shown in section 3 indicate that the sum of all these other ion sources/sinks as well as ion flows into the divertor region which also flow to the outer target divertor appear either to be negligible or to balance each other.

## 4.4 Applicability of TCV results to other existing and planned tokamaks

A central focus of this paper is the development of target ion current loss in detachment which is set in motion when the power flowing into the recycling region drops to twice the level required for the ion source - $q_{recl}/(E_{ion} \Gamma_t)$ ~ 2. This leads to a 'power limitation' of the ion source. This appears to be the main driver of the $I_t$ roll-over on TCV, while recombination has a much smaller effect and occurs after the roll-over of $I_t$. Power limitation can play a dominant role in ion current loss during detachment when $I_t$ is (almost) fully delivered by the divertor ion source (e.g. 'high recycling'), which is verified in this paper for TCV. The assumption that the divertor is in a high-recycling condition is more likely true for closed, higher density, divertors than for open divertors such as TCV. That said, preliminary results at lower current between both field (B) directions indicate that the magnitude of possible ion flows into the divertor ($I_{SOL}$ – section 4.3) may be changed with the field direction. Such changes are likely due to drift effects [80]. Additionally, the power asymmetry changes with field direction [68]. This may influence to what degree ionisation contributes to the ion target current. However, measurement accuracy is reduced in these conditions due to the low divertor densities reducing the Stark broadening density inference accuracy.

A second question regarding the wider applicability of the TCV results is on the timing (and magnitude) of the significant contribution of the electron-ion recombination sink. In TCV, this is both after the detachment onset ($P_{ion}/P_{recl}$ ~ 0.5; $T_t$ ~ 4-7 eV) and the target ion current roll-over ($P_{ion}/P_{recl}$ ~ 0.7; $T_t$ ~ 2 eV). Instead, volumetric recombination becomes significant at temperatures < 1 eV, which occurs when $P_{ion}$ ~ $P_{recl}$. Therefore, it seems that power limitation and the detachment onset ($P_{ion}/P_{recl}$ ~ 0.5) occur *before* volumetric recombination becomes significant, which is expected to be general even to higher density and higher power machines as the argument based on temperature. 'Power limitation' ($q_{recl}/(E_{ion} \Gamma_t) < 2$) is thus expected to be a requirement in high recycling regimes to reach the conditions for limiting the ion source and for lowering the target temperature to reach conditions for significant volumetric momentum loss and then finally, recombination. This is supported by our quantitative results and qualitative estimates on C-Mod [8]; analytic modelling as well as SOLPS modelling for TCV [35] and that for other devices [12, 81].

However, how 'quickly' one goes from $P_{ion}/P_{recl}$ ~ 0.5 to $P_{ion}/P_{recl}$ ~ 1 (and thus recombination relevant conditions) during detachment can depend on a range of parameters (including how quickly momentum loss develops [1]) and is likely better addressed in fluid models of higher density plasmas.

When $P_{ion}/P_{recl}$ ~ 1 conditions are achieved, the importance of volumetric recombination on TCV is significantly smaller than in higher density devices, such as C-Mod where volumetric recombination can drop the ion target flux by a factor 10-100 during a core density ramp [8]. This could also result in a more significant movement of the recombination and density peaks (front) at the deepest detached conditions.

We note that the effect of $N_2$-seeding to reach detachment strongly reduces the level and importance of recombination as an ion sink for TCV. This does appear to scale to higher density tokamaks, such as C-Mod [8] and JET [32]. It thus seems generally true that volumetric recombination is not a requirement for (roll-over) detachment.

We do expect the characteristic gradient scale lengths of various quantities such as ionization, recombination and CX to be shorter (poloidal and along B) in tokamaks with higher densities and parallel power densities than for TCV. Certainly the parallel heat flux would be 100x larger in ITER than TCV leading to smaller parallel-to-B temperature scale lengths in absolute value and relative to the divertor size $\Delta L_{q\|}$ ($\Delta L_{q\|} \sim \Delta T / q_\|$ where $\Delta T$ is set by the impurity cooling curve, ~ 10s of eV for carbon) [82]; this would lead to more localized impurity radiation and ionisation regions than in TCV. In addition, higher density will lead to shorter CX and ionization mean free paths. Higher divertor densities, for the same upstream conditions, may be facilitated by the planned baffle upgrade [83] of TCV.

Operation in H-mode further shortens those gradients and leads to even more localised impurity radiation and ionisation regions. Furthermore, it affects particle and power fluxes and may also change the flow of the impurities in the divertor, which would impact momentum losses.

There is another likely key change in divertor characteristics engendered by larger $P_{sep}$ and $q_{||}$. Intrinsic carbon radiation in TCV suffices to lower $q_{recl}$ so that it limits the ionization source during density ramp discharges. However, as $q_{||}$ is increased, reaching $P_{recl} \sim 2 \times P_{ion}$ without additional impurity seeding is correspondingly more difficult to accomplish during density ramps only [84]. That is particularly true for operation with high-Z metallic walls where we expect less intrinsic divertor radiation, adding impetus to needing seeded impurities to detach. However, given that impurity seeded TCV plasmas clearly show lower volumetric recombination (also true for JET [32] and C-Mod [8]) than for density ramp-driven detachment, the connection between seeding and recombination needs to be better understood.

In addition to the discussion above, several caveats for general divertor detachment investigations have emerged from this study, relevant for detachment investigations on other devices. First, the linear increase of the target ion flux on TCV with upstream density for an attached divertor is consistent with considering all aspects of the 2PMR model. This points out that the often-used degree of detachment (DoD) scaling ($\Gamma_t \propto \overline{n_e}^2$) must be modified to account for changes in upstream parameters ($n_e$, $T_e$) and divertor radiation. Lastly, target pressure loss during detachment can be due to both volumetric momentum losses and a drop in the upstream pressure; it is unclear whether the upstream pressure loss is driven by upstream processes (e.g. cross-field transport) or by changes in the divertor, or both.

## 5. Summary

Spectroscopic measurements of the TCV outer divertor plasma, combined with novel analysis techniques, has enabled an in-depth study of the roles of various processes (ion and power sources and sinks) controlling the divertor target ion current during detachment. Of particular importance to this study is the new ability to determine the poloidal divertor ionization source profile, and thus the total divertor ion source. These novel measurements provide the *first experimental verification that the ion source ($I_i$) in the divertor can be the primary determinant of the target ion current ($I_t$) from attached conditions through the detachment onset and $I_t$ drop (roll-over)* in TCV. The volumetric electron-ion recombination ion sink is relatively small or negligible until after the roll-over of $I_t$ when $T_t$ reaches low values. *Volumetric recombination thus seems not to be a requirement for detachment and should only occur at temperatures lower than when the ionization source is limited.*

Our power balance measurements during a core density ramp show that the onset of detachment occurs at a point when the power flowing into the divertor minus divertor impurity radiation (the power flowing into the recycling region), $P_{recl}$, drops to a value that is twice $P_{ion}$, the measured power required for ionization ($f_{ion} \equiv P_{recl}/P_{ion} \sim 2$). At that point, the target temperature $T_t \sim 4$-7 eV and the target ion current deviates from the expected attached scaling (linearly with upstream density on TCV). As $P_{recl}/P_{ion}$ and $T_t$ continue to drop during a core density scan, the ion source and target ion current start dropping (roll-over) at $P_{recl}/P_{ion} \sim 1.4$ ($T_t \sim 2$ eV). As $P_{recl}/P_{ion}$ approaches 1, where little thermal/kinetic power reaches the target ($T_t \leq 1$ eV), the ion-electron recombination sink for ions can become significant, but only after the divertor ionization source is limited by $P_{recl}$. *The above sequence, as well* *as power reduction to the ionization and target regions driving the detachment process, are expected to be general beyond TCV.* Essentially the same sequence is followed when using impurity seeding to reach detachment where we find that the role of recombination is further diminished.

We have also shown that our experimental measurements are consistent with analytic and 2D modelling predictions. *Simple power and particle balance analytic models, using target temperature measurements, predict the target ion current from attached through detachment onset and the current roll-over in quantitative agreement with the $I_t$ measurement.* It also shows that the ion source can be written as a trade-off between the maximum possible ion source ($P_{recl}/E_{ion}$) and fraction of that power spent on ionisation ($f_{ion} = P_{ion}/P_{recl}$), which increases with decreasing $T_t/E_{ion}$. The $f_{ion}$ predicted from $T_t/E_{ion}$ quantitatively agrees with $f_{ion}$ obtained directly from spectroscopic measurements.

However, the ion source prediction from power and particle balance must also be consistent with the sheath conditions ($p_t \propto \Gamma_t T_t^{1/2}$) – as is done in the '2PMR' model in this work. That consistency leads to three equivalent quantitative predictions for the detachment onset: $T_t = E_{ion}/\gamma \sim$ 4-7 eV, $f_{ion}$= 0.5, and $p_t/q_{recl} = \sqrt{\frac{m_i}{2E_{ion}\gamma}} \sim$ 8 N/MW. *All three have been found to match, within uncertainties, the experimentally-determined detachment onset.* The extension of the $I_t$ prediction beyond these thresholds *requires* pressure loss. The observation that *atomic physics* supplies volumetric momentum losses when the detachment onset criteria requires it (when required by plasma physics) is striking.

Our measurements have further validated the physics included in the SOLPS modelling code. *Measured outer divertor poloidal profiles of ion/power sources, sinks and other plasma parameters are compared with SOLPS predictions for three points in the detachment process with generally good quantitative agreement.*

# Acknowledgments

The authors would like to thank B. Dudson, S. Krasheninnikov, D. Moulton and P. Stangeby, for discussions of the physics processes, previous work and spectroscopic techniques relevant to this paper. This work has been carried out within the framework of the EUROfusion Consortium and has received funding from the Euratom research and training programme 2019–2020 under Grant Agreement No. 633053. The views and opinions expressed herein do not necessarily reflect those of the European Commission. This work was supported in part by the Swiss National Science Foundation. This work was supported in part by the US Department of Energy under Award Number DE-SC0010529. The PhD research of K. Verhaegh was supported by funding from the University of York and the Swiss National Science Foundation. B. Lipschultz was funded in part by the Wolfson Foundation and UK Royal Society through a Royal Society Wolfson Research Merit Award as well as by the RCUK Energy Programme (EPSRC grant number EP/I501045).

## A.1 Two point model with recycling energy losses (2PMR)

In literature [4, 11], the effect of recycling losses has been added to the Two Point model [4], which we refer to as the "2PMR". In this section, a more explicit derivation of adding the effect of recycling energy losses to the Two Point model is provided; which has been utilised for several predictions in section 4.2. This 2PMR model provides both a quantitative criterion for the expected onset of detachment as well as predictions of the ion target current/target temperature given measurements of $q_{recl}$, $E_{ion}$ and $p_t$ (or, assuming a known $f_{mom}$ of momentum losses, $p_t = f_{mom} \; p_u$). A more detailed discussion on this '2PMR' can be found in [1].

## A.1.1 2PMR derivation

We assume the target ion flux ($\Gamma_t$), on a single flux tube, is fully determined by the ionisation source on that particular flux tube ($\Gamma_i$) (Equation A.1). With that, it is implicitly assumed that both ion flows from outside of the ionisation region towards the target as well as volumetric recombination in the divertor are negligible. Furthermore, cross-field transport of heat and/or particles is ignored.

$$\Gamma_t = \Gamma_i \tag{A.1}$$

It is defined that $q_{recl}$ equals the kinetic part of the heat flux reaching the target $q_t = \gamma \Gamma_i T_t$ plus the power flux spent on ionisation $E_{ion}\Gamma_i$, yielding a relation relating $q_{recl}$ to $q_t$, and $T_t^*$ which is defined as the ratio between energy spent on ionisation and kinetic energy reaching the target $T_t^* = \frac{\gamma T_t}{E_{ion}}$, as explained in section 4.1. When thinking of the divertor as separate ionisation/recycling and impurity radiation regions, $q_{recl}$ physically represents the power flux "entering" the recycling region. However, our arguments do not depend on the localisation or possible overlap of these regions.

$$q_{recl} = E_{ion}\Gamma_i + q_t = \gamma \Gamma_i T_t \left(1 + \frac{E_{ion}}{\gamma T_t}\right)$$

$$\rightarrow q_t = q_{recl}\frac{T_t^*}{1+T_t^*} = q_{recl}\, f_{kin} \tag{A.2}$$

Using the sheath target conditions (Equation A.3), a relation for the target temperature as function of the heat flux reaching the target and the target pressure $p_t$ can be established (note that $p_t$ is the total target pressure); which is similar to the "default" two point model result [4]. Here it is assumed that the Mach velocity near the target is 1. We deliberately utilise the target pressure $p_t$, instead of the upstream pressure $p_u$ in equation A.3 to make the derivation more general

$$q_t = \gamma n_t T_t \sqrt{\frac{2T_t}{m_i}}$$

$$\rightarrow T_t = \frac{2\, m_i}{\gamma^2}\left(\frac{q_t}{p_t}\right)^2 \tag{A.3}$$

By combining Equation A.3 with Equation A.2 a prediction for $T_t$ is obtained (Equation A.4.1), which is the central equation of the 2PMR. This equation can be re-written in a quadratic form (equation A.4.2)

$$T_t = \frac{2\, m_i}{\gamma^2}\left(\frac{q_{recl}}{p_t}\right)^2 \left(\frac{\gamma T_t}{\gamma T_t + E_{ion}}\right)^2 \tag{A.4.1}$$

$$T_t^2 + \left(\frac{2E_{ion}}{\gamma} - \frac{2m_i}{\gamma^2}\left(\frac{q_{recl}}{p_t}\right)^2\right)T_t + \frac{E_{ion}^2}{\gamma^2} = 0 \tag{A.4.2}$$

Before discussing solutions of equation A.4, which provide $T_t$ as function of $q_{recl}$, $p_t$ and $E_{ion}$, first we will utilise equation A.4.1 to derive two expressions for $\Gamma_t$ from the 2PMR.

First, we combine Equation A.2 and Equation A.4.1 to obtain a relation for the ion target current flux in terms of $p_t$, $q_{recl}$, $E_{ion}$ and physical constants – equation A.5. This equation has an equivalent form to the 'default' Two Point Model expected ion target current trend (e.g. $\Gamma_t = \frac{\gamma\, p_t^2}{2\, m_i\, q_t}$ where $q_t = q_{recl}\, f_{kin}$), which emphasizes the role of target pressure loss during ion target current loss as emphasized by equation 2. Here $f_{kin}$ ($T_t^*$) is identical to $f_{kin}$ introduced in section 4.1.

$$q_{recl} = E_{ion}\Gamma_i + q_t = \gamma \Gamma_t T_t \left(1 + \frac{E_{ion}}{\gamma T_t}\right)$$

$$\Gamma_t = \frac{q_{recl}}{\gamma T_t + E_{ion}} = \frac{q_{recl}}{\gamma \frac{2\,m_i}{\gamma^2}\left(\frac{q_{recl}}{p_t}\right)^2 \left(\frac{\gamma T_t}{\gamma T_t + E_{ion}}\right)^2} \frac{\gamma T_t}{E_{ion} + \gamma T_t}$$

$$\Gamma_t = \frac{\gamma\, p_t^2}{2\, m_i\, q_{recl}\, f_{kin}(T_t^*)} \tag{A.5}$$

It is important to note that Equation A.5 is identical to Equation A.6 which is essentially the flux surface's equivalent of Equation 18. Here $f_{ion}(T_t^*)$ is identical to $f_{ion}$ introduced in section 4.1. Equations A.5, A.6 can be evaluated using the expressions for $f_{ion}$, $f_{kin}$ with the $T_t$ result obtained by solving equation A.4.1.

$$q_{recl} = E_{ion}\Gamma_i + q_t = \gamma \Gamma_t T_t \left(1 + \frac{E_{ion}}{\gamma T_t}\right)$$

$$\Gamma_t = \frac{q_{recl}}{\gamma T_t + E_{ion}} = \frac{q_{recl}}{E_{ion}} \frac{E_{ion}}{E_{ion} + \gamma T_t}$$

$$\Gamma_t = \frac{q_{recl}}{E_{ion}} f_{ion}(T_t^*) \tag{A.6}$$

Therefore, the 2PMR essentially provides a bridge between the power/particle balance model treated in section 4.1, and the 'default' Two Point model for divertor modelling, which emphasizes the role of momentum balance. It also shows that an $\Gamma_t$ prediction from power/particle balance (equation A.6) accounts for momentum balance implicitly through $T_t$ (which depends on momentum balance/losses). Similarly, an $\Gamma_t$ prediction from momentum balance (equation A.5) accounts for power/particle balance implicitly through including the target heat flux.

### A.1.2 Solving for $T_t$ in the 2PMR

Now that we discussed the implications of the equivalence of considering the ion target current from power/particle and momentum balance point of views, we will discuss solutions to equation A.4 to obtain a target temperature estimate from the 2PMR using $p_t/q_{recl}$, $E_{ion}$.

Equation A.4 can be solved either numerically or, under the assumption that $p_t/q_{recl}$ (or $p_u/q_{recl}$ with $p_u=p_t f_{mom}$) and $E_{ion}$ are independent/control variables to determine $T_t^*$, analytically as a quadratic equation, resulting in Equation A.7. Note that this *explicitly implies that $p_t/q_{recl}$ (or $p_u/q_{recl}$ with $p_u=p_t f_{mom}$) as well as $E_{ion}$ do not have an additional target temperature dependence apart from equation A.4*. In that case equation A.4.2 is not a quadratic equation anymore and cannot be solved as such. Such assumptions are commonly used in analytic divertor models [4, 9, 11, 15] and likely apply in attached conditions.

$$T_t = \left(\frac{m_i}{\gamma^2}\left(\frac{q_{recl}}{p_t}\right)^2 - \frac{E_{ion}}{\gamma}\right) \pm \sqrt{\frac{m_i}{\gamma^2}\left(\frac{q_{recl}}{p_t}\right)} \sqrt{\frac{m_i}{\gamma^2}\left(\frac{q_{recl}}{p_t}\right)^2 - \frac{2 E_{ion}}{\gamma}} \tag{A.7}$$

The quadratic equation for $T_t$ has two solutions (Equation A.7), of which only the positive branch ($T_t \geq \frac{E_{ion}}{\gamma}$;) is stable (Equation A.8) in steady-state conditions as has been explained in literature [4, 9, 11]. Furthermore, 1D time-dependent detachment simulations have been studied using SD1D in conditions indicative to those analytically described by the negative branch of equation A.7 [71]. In these simulations, $p_u$ is fixed and momentum losses are removed (e.g. thus fixing $p_t$), resulting in oscillations and supersonic flows when accessing $T_t < \frac{E_{ion}}{\gamma}$, which may be unrealistic [71]. Hence, we assume that the negative branch, $T_t \leq \frac{E_{ion}}{\gamma}$, of Equation A.5 cannot occur.

In the analytic model we see that the target temperature drops when $\left(\frac{q_{recl}}{p_t}\right)$ decreases, which is consistent with experimental conditions where a density ramp or impurity seeding (e.g. a drop in $\frac{q_{recl}}{p_t}$) results in a target temperature reduction. This temperature drop occurs until $\frac{m_i}{\gamma^2}\left(\frac{q_{recl}}{p_t}\right)^2 - \frac{2E_{ion}}{\gamma} = 0$ is reached (equation A.7), at which the minimum temperature possible in the positive branch of equation A.7 is reached: $T_t = \frac{E_{ion}}{\gamma} \sim 4$ eV (for $E_{ion}$=28 eV and $\gamma$=7). Although $E_{ion}$ increases as the divertor cools, this increase of $E_{ion}$ is negligible until this temperature (see section 4.1.5 and [1]). When combining the temperature expression in equation A.7 with the ion target current predictions (equation A.5 and A.6), a decrease in $\frac{q_{recl}}{p_t}$ (where $T_t$ decreases up until $T_t = \frac{E_{ion}}{\gamma}$) will *not* result in an ion current roll-over.

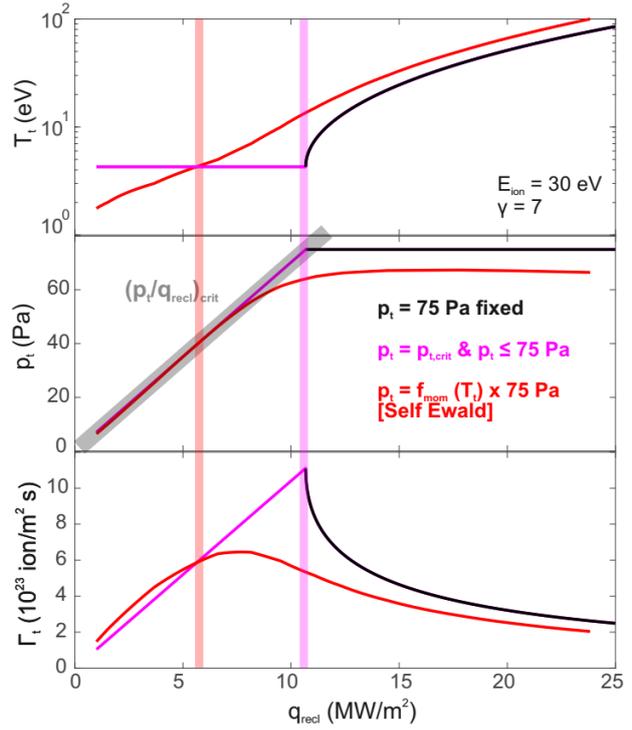

*Figure A.1: 2PMR results of $T_t$, $p_t$, $\Gamma_t$ as function of $q_{rec}$ for three different target pressure behaviours during detachment. The analytic predicted detachment onset (equation A.9) is shown with the vertical bars. The maximum $p_t$ at $(p_t/q_{recl})_{crit}$ is shown.*

All of this implies that the equation A.7 *cannot* be used to model the detached state, which has temperatures lower than 4 eV and an ion current roll-over. An example of this is shown in figure A.1, where $p_t$, $E_{ion}$ and $q_{recl}$ are independent parameters, of which $p_t$, $E_{ion}$ are fixed and $q_{recl}$ is scanned. $T_t$ and $\Gamma_t$ are shown as function of $q_{recl}$, providing a solution for $T_t \geq \frac{E_{ion}}{\gamma}$.

The above remains true even if a *fixed* (e.g. not explicitly dependent on $T_t$) momentum loss through $p_t$ = $f_{mom}$ $p_u$ is introduced. This would reduce $p_t/q_{recl}$, reducing the ion target current while *also increasing* the target temperature. This raises the question, what is required to get a detachment-like behaviour in the 2PMR?

### A.1.3 2PMR and detachment

A simultaneous drop of both $\Gamma_t$ and $T_t$ implies $\frac{\partial}{\partial T_t}\Gamma_t > 0$, which requires, according to equation 2: $\frac{\partial}{\partial T_t}\Gamma_t \propto \frac{\partial}{\partial T_t}(p_t T_t^{-\frac{1}{2}}) > 0$. That equation can only hold if the target pressure is dropping as a *specific function* of the target temperature and obeys: $\frac{\partial p_t}{\partial T_t} > \frac{p_t}{2\,T_t}$. This implies, for instance, that when parametrising $p_t(T_t^*) = p_0 T_t^{*\alpha}$ in the detached ($T_t^* < 1$) regime, roll-over only occurs for $\alpha > \frac{1}{2}$ [1]. This target pressure loss could either be due to momentum losses increasing with target temperature or due to the upstream pressure dropping with target temperature, or a combination of both.

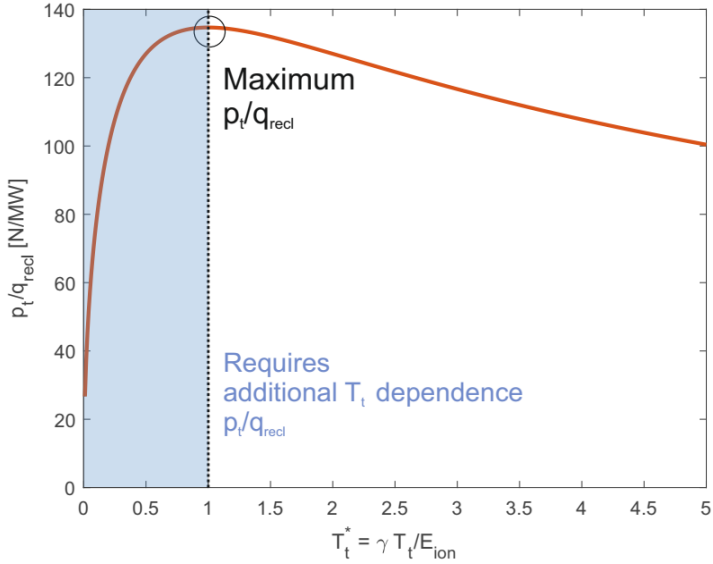

Figure A.2: Visualisation of $\frac{p_t}{q_{recl}} = \gamma\sqrt{2\,m_i}E_{ion}^{-1/2}T_t^{*1/2}\frac{1}{T_t^*+1}$ as function of $T_t$ for deuterium with $E_{ion}$ = 28 eV and $\gamma = 7$. The regions where the maximum $p_t/q_{recl}$ value is achieved and where a reduction of $p_t/q_{recl}$ as function of $T_t$ is required are highlighted.

Parametrisations in literature for $p_t$ are, for instance, obtained by assuming that the upstream pressure is constant and assuming a parametrised form of $f_{mom}$ as function of the target temperature. Those forms of $f_{mom}$ are, for instance, retrieved by the Self-Ewald model [9] or by analytic fits through measurements/SOLPS data [9, 85].

However, such parametrisation functions are likely experiment/machine specific, are not easily experimentally obtained and their determination is generally ad-hoc. Therefore, it is preferable to analyse the 2PMR without relying on this parametrisation.

To do this, we express the general solution to equation 4.1 graphically in Figure A.2 in the form of $\frac{p_t}{q_{recl}} = \gamma\sqrt{2\,m_i}E_{ion}^{-1/2}T_t^{*1/2}\frac{1}{T_t^*+1}$ (where $T_t^* = \frac{T_t\gamma}{E_{ion}}$). The intersection of $p_t$ ($T_t^*$) curve with the shown solution would provide the solution to $T_t^*$. Figure A.2 is thus generally applicable, regardless of the parametrisation $p_t$ ($T_t^*$).

Considering that relation, $p_t/q_{recl}$ increases as $T_t$ is decreased until a maximum value $\left(\frac{p_t}{q_{recl}}\right) = \sqrt{\frac{m_i}{2E_{ion}\gamma}}$ is reached after which $\left(\frac{p_t}{q_{recl}}\right)$ has to drop continuously. This implies that there is a *maximum possible target pressure*, given a value for $q_{recl}$ and $E_{ion}$. This also means that, given a ratio between the target and upstream pressure ($f_{mom} \equiv p_t/p_u$), there is a *maximum possible upstream pressure*, expressed by equation A.9. That maximum target pressure is reached at $T_t = \frac{E_{ion}}{\gamma}$ and thus $f_{ion} = f_{kin} = 0.5$. Therefore, to reduce the target temperature below that value, a reduction of $\frac{p_t}{q_{recl}}$ is required and the magnitude of the reduction increases with lower target temperatures.

$$\left(\frac{p_u}{q_{recl}}\right)_{crit} = \frac{1}{f_{mom}}\sqrt{\frac{m_i}{2E_{ion}\gamma}} \tag{A.8}$$

The *critical point* where *the start of* such a $\frac{p_t}{q_{recl}}$ reduction is *required* can be written as three equivalent criteria – equation A.9. The required reduction increases when considering $E_{ion}$ increases during detachment [1]. Note that, if one were to implement a parametrisation of $p_t$ ($T_t$), the target pressure can drop *earlier* than this point. Hence, these critical points physically represent the *lowest possible temperature at which target pressure loss must occur*.

$$\left(\frac{p_t}{q_{recl}}\right)_{crit} = \sqrt{\frac{m_i}{2E_{ion}\gamma}} = \frac{1}{\gamma\, c_{st}\left(T_t=\frac{E_{ion}}{\gamma}\right)} \qquad (A.9.1)^2$$

$$T_{t,crit} = \left(\frac{E_{ion}}{\gamma}\right) \qquad (A.9.2)$$

$$f_{ion,crit} = 1/2 \qquad (A.9.3)$$

This point has been defined by Krasheninnikov [11, 15] as the onset of detachment and by Stangeby [4, 9] as the limit of a regime where solutions such as equation A.7 stop applying and momentum losses need to be accounted. We use these criteria of the *lowest possible temperature at which target pressure loss must occur* as an approximation of the *onset of target pressure loss*, which we define as the detachment onset. The onset of target pressure loss is important detachment since this implies that $\Gamma_t\, T_t^{-1/2}$ starts to drop – equation 2, which is required before detachment (e.g. both $\Gamma_t$ and $T_t$ drop simultaneously) can occur. This agrees with the experimental findings in the paper.

However, target pressure loss could – theoretically - occur *before* it is *required*. Since the ion target current roll-over occurs when $\frac{\partial}{\partial T_t}\Gamma_t \propto \frac{\partial}{\partial T_t}(p_t T_t^{-\frac{1}{2}}) > 0$, this could occur before the detachment onset criteria if $p_t$ drops sufficiently rapidly before the detachment onset. However, experimentally in this paper we show that this does *not* occur on TCV.

Now we can ask the question on how much target pressure loss is *required* during detachment? Certainly, in a case where $q_{recl}$, $p_t$, $E_{ion}$ are free parameters where $q_{recl}$ is scanned, we obtain a point where the maximum ratio $p_t/q_{recl}$ (equation A.9.1) is reached, as shown in figure A.1. Since this ratio cannot go any higher, any further reduction of $q_{recl}$ *requires at least* an equal reduction in $p_t$ such that $p_t/q_{recl}$ does not exceed the maximum possible value – e.g. stays at the peak shown in figure A.1. In other words, this provides us with the *minimum required reduction* of $p_t$ when $q_{recl}$ is further decreased as $\left(\frac{p_t}{q_{recl}}\right)$ remains at its *maximum possible* level (e.g. $\sqrt{\frac{m_i}{2E_{ion}\gamma}}$). In that case the target temperature would remain stuck at $T_{t,crit} = \left(\frac{E_{ion}}{\gamma}\right)$ (equation A.4).

What happens when we obtain a *larger* reduction of the target pressure than this minimum reduction? A *larger* reduction of $\frac{p_t}{q_{recl}}$ (likely through a target pressure reduction) *than this minimum* enables lower target temperatures, as shown in figure A.1. This requires the target pressure *to become parametrised function of the target temperature* [4, 9], which is also required for a *simultaneous* drop of $T_t$ and $\Gamma_t$ as discussed previously. Again, an example of this is shown in figure A.1 where the Self-Ewald model was used to obtain a $f_{mom}$ ($T_t$) is dependence while a fixed value for $p_u$ is assumed (e.g. $p_t$ ($T_t$) = $f_{mom}$ ($T_t$) $p_u$). The result shows that both $p_t$ and $\Gamma_t$ drop as $q_{recl}$ is decreased beyond the detachment onset. With a $p_t$ reduction given by the Self-Ewald model, $p_t$ already drops at higher temperatures *before* the point where a $p_t$ drop is *required*. Therefore, a small deviation occurs between the detachment onset prediction (which corresponds to the *minimum* $T_t$ at which $p_t$ needs to start to drop) and the ion current roll-over. This depends strongly on the parametrisation of $p_t$ ($T_t$), which as shown in [1], influences how quickly $T_t$, $p_t$ and $\Gamma_t$ drop in the detached case as function of $q_{recl}$.

---

[2] This critical threshold (in the form $p_u/q_{recl}$ – this will be explained later) is twice larger than in [9]. However, the calculated threshold (Figure 15c) is ~2 x smaller than the quoted 15 N/MW [9] likely due to different values used for $E_{ion}, \gamma$. However, this point is identical to the $p_u^2 / q_{||}$ reached in [2,7], using Equations 5.37, 5.39 on p. 237, 238 with $T_t = E_{ion} / \gamma$.

## A.2 Evaluating and applying the 2PMR model with experimental data

We will now explain how we will utilise the analytic models in section A.1 to apply them to our experimental measurements. First, we aim to predict the *detachment onset for the flux surface corresponding to the separatrix*. This means that $p_t$, $E_{ion}$ and $q_{recl}$ should correspond to their separatrix values. Since no $p_t$ values are available, we assume pressure balance $p_t = p_u$, which seems experimentally an accurate assumption before the detachment onset. Although $p_u$ can be obtained at the separatrix (from Thomson scattering from the chord closest to the separatrix), assumptions must be made to estimate $E_{ion}$ and $q_{recl}$ on the separatrix.

As explained in sections 4.1.3 and [2], an estimate of $E_{ion}$ is obtained from spectroscopic inferences, which provides an "effective" $E_{ion}$, which is divertor averaged over all the different flux surfaces. Assuming this is the same as $E_{ion}$ at the separatrix has a negligible effect on the 2PMR predicted detachment onset.

To estimate $q_{recycl}$ at the separatrix we divide the power entering the ionisation region ($P_{recl}$ – section 3.3), with an effective area, $A_{eff} = 2\pi R_{target} \frac{B_t}{B_p} \lambda_{SOL}$ [4], where it is assumed that the radial location of the ionisation region is the same as the target radius. $A_{eff}$ depends on the ratio between the toroidal and poloidal field ($\frac{B_t}{B_p}$) and scrape-off-layer width $\lambda_{SOL}$. The SOL width has been approximated by using $\lambda_{q,int}$ of the heat flux profile measured through IR imaging at the target, which has been mapped upstream [44]. The choice for a characterisation using $\lambda_{q,int}$ for the spatial profile of $q_{recl}$ across flux surfaces has been made as this parametrisation is more robust during detached regimes than the Eich fit [44]. It is assumed that the spatial profile of $q_{recl}$ is the same as that of the target heat flux, which enables extracting $q_{recl}$ from $P_{recl}$ using $\lambda_{q,int}$. Volumetric radiation could, however, alter this heat flux shape. This shape modification is, however, expected to be small as most of the radiative dissipation happens in the impurity radiation region upstream of both the target and the recycling region. Uncertainties of the characterization of $A_{eff}$ have been neglected and could lead to systematic deviations from the portrayed trend of $q_{recl}$.

Instead of estimating the detachment onset, we also wish to apply this technique to model the behaviour of *the integrated ion current as function of 'upstream' parameters ($p_u$, $q_{recl}$)*, which can be compared with the experimentally measured integrated ion current. This requires Equation A.5 (or A.6) to be integrated along the entire divertor floor – Equation A.10, where $n_u$ is the upstream density and $T_u$ is the upstream temperature. For this we utilise the $T_t$ expression from equation A.7, where we assume $p_t = p_u$ is a control parameter. For the sake of simplicity, we assume that $f_{kin}$ at the separatrix is characteristic for the entire divertor. Such an assumption can likely be made since the influence of $f_{kin}$ on Equation A.5 is limited as $f_{kin}$ can only vary between 0.5 and 1.

$$I_t = \int 2\pi r\, \Gamma_t\, dr$$

$$I_t = \frac{\gamma \pi}{m_i} \frac{1}{f_{kin}} \times \int r\, \frac{n_u(r)^2 T_u(r)^2}{q_{recl}(r)} dr \quad \text{(A.10)}$$

To simplify the expression of the integral, the upstream density and temperature profiles are broken up in their separatrix values (e.g. $n_u^0$) times a function describing their profile behaviour (e.g. $f_{n_u}(r)$) as shown in Equation A.11.

$$n_u(r) = f_{n_u}(r) n_u^0$$

$$T_u(r) = f_{T_u}(r) T_u^0$$

$$q_{recl}(r) = f_{q_{recl}}(r) T_u^0 \quad \text{(A.11)}$$

To maximize temporal resolution, the upstream separatrix density/temperature are obtained from Thomson scattering, while the normalised upstream density/temperature profiles ($f_{n,u}(r)$, $f_{T,u}(r)$) are obtained by fitting reciprocating probe upstream density/temperature profiles, at the probe plunge times, with a double exponential: $f_{n_u} = A_1 e^{-\frac{r-r_{rsep}}{\lambda_1}} + A_2 e^{-\frac{r-r_{rsep}}{\lambda_2}}$ where $r_{sep}$ is the separatrix radius upstream. A single exponential profile using $\lambda_{q,int}$ has been used to describe the profile of $q_{recl}(r)$, whose integral equals $P_{recl}$ as shown in equation A.12, again assuming that the heat flux shape at the target is similar to the heat flux shape of $q_{recl}$ upstream entering the recycling region. Since the model of appendix A.1 requires the parallel heat flux ($q_{recl}^{\parallel}$) equation A.12 also shows the field mapping used.

$$f_{q_{recl}}(r) = e^{-\frac{r-R_{sep}}{\lambda_{q,int}}}$$

$$q_{recl}^0 = \frac{\int 2\pi r\, f_{q_{recl}}(r)\, dr}{P_{recl}} \approx \frac{2\pi R_{sep} \lambda_{q,int}}{P_{recl}}$$

$$q_{recl}^{\parallel} = \frac{B_p}{B_t} q_{recl} \qquad \text{A.12}$$

Using these profile expressions, the target ion flux can be expressed as shown in Equation A.13. The modelled integrated target ion current scales as $\frac{1}{f_{kin}}$ (evaluated at the separatrix), times $\frac{p_u^{0^2}}{P_{recl}}$ (where $p_u^0$ is evaluated at the separatrix) times $f_p$, which is a parameter describing the influence of the evolution of all spatial profiles as well as divertor geometry on $I_t$ as indicated in Equation A.13, which is integrated from the separatrix until infinity.

$f_p$ outside of the reciprocating probe plunge times is interpolated by fitting a polynomial to $f_p$ across all probe plunge times. Uncertainties in $p_u^0$, $P_{recl}$ and $f_{kin}$ are accounted for, while uncertainties in the profile description are neglected. The separatrix values of the upstream density/temperature/target temperature are referred to as $n_u$, $T_u$, $T_t$ in other parts of the paper.

$$I_t = \frac{2\gamma \pi^2}{m_i} \times \frac{1}{f_{kin}} \times f_p \times \frac{p_u^{0^2}}{P_{recl}}$$

$$f_p = R_{sep} \lambda_{q,int} \frac{B_t}{B_p} \int_{r_{sep}}^{\infty} r\, \frac{f_{n_u}^2(r) f_{T_u}^2(r)}{f_{q_{recl}}(r)}\, dr \qquad \text{(A.13)}$$